\documentclass[trackchanges, twocolumn, tighten, times, twocolappendix]{aastex631}

\usepackage{subfigure}
\usepackage{comment}
\usepackage{amsmath}
\usepackage{bm}
\usepackage{enumitem}
\usepackage{hyperref}

\begin{document}

\vspace*{-1.25cm}

\title{Studying the mirror acceleration via kinetic simulations of relativistic plasma turbulence}

\author[0000-0001-5796-225X]{Saikat Das}
\email{saikatdas@ufl.edu}
\affiliation{Department of Physics, University of Florida, Gainesville, FL 32611, USA}

\author[0000-0002-0458-7828]{Siyao Xu}
\email{xusiyao@ufl.edu}
\affiliation{Department of Physics, University of Florida, Gainesville, FL 32611, USA}

\author[0000-0002-3226-4575]{Joonas N\"{a}ttil\"{a}}
\email{joonas.nattila@helsinki.fi}
\affiliation{Department of Physics, University of Helsinki, P.O. Box 64, University of Helsinki, FI-00014, Finland}

\begin{abstract}

Efficient relativistic turbulent acceleration of particles is indicated by recent astrophysical observations. 
{The acceleration mechanism due to temporal variations of magnetic field strengths (``Type II mechanism")} 
remains underexplored. 
The mirror acceleration has recently been proposed as an efficient Type II mechanism for particle energization in turbulence-compressed magnetic fields.  
We perform a 3D particle-in-cell (PIC) simulation of pair plasma  
to extend its study to relativistic turbulence. 
By tracking individual particles, we see 
that the particles interacting with transverse magnetic mirrors can have a 
significant energy gain during one mirror interaction and within one gyro-orbit. 
As expected for the mirror acceleration, we statistically find that the momentum 
gain is preferentially in the direction perpendicular to the local magnetic field and 
positively correlated with the local magnetic field strengthening.
As a result, 
the particle pitch angle distribution becomes increasingly anisotropic toward higher energies, with a concentration at large pitch angles.  
The mirror acceleration facilitates a spatial confinement of particles by stochastically increasing their pitch angles, which further enhances the 
mirror acceleration.

\end{abstract}




\section{Introduction} \label{sec:intro}

\begin{figure*}
    \subfigure[]{
    \includegraphics[width = 0.33\textwidth]{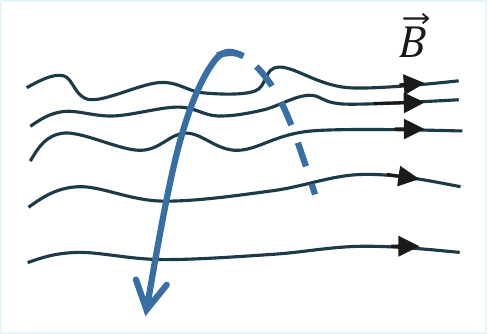}
    }%
    \hfill
    \subfigure[]{
    \includegraphics[width = 0.33\textwidth]
    {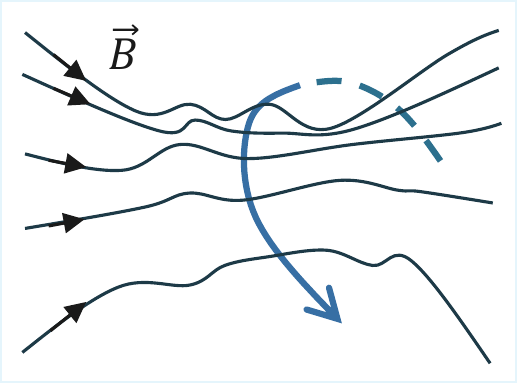}
    }%
    \hfill
    \subfigure[]{
    \includegraphics[width = 0.24\textwidth]{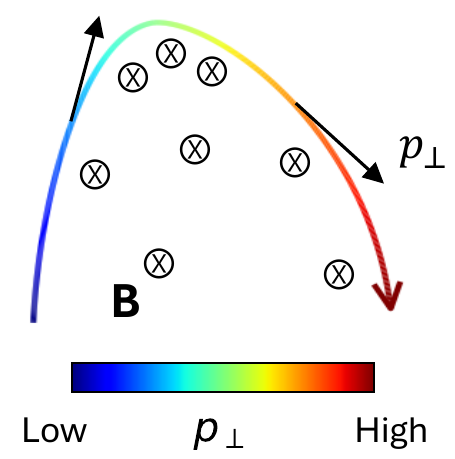}
    }
    \caption{(a) Interaction of a particle with a transverse magnetic mirror, where the blue line indicates the particle trajectory, and the black lines indicate the magnetic field. 
    (b) Interaction of a particle 
    with both a transverse mirror and a longitudinal mirror. 
    (c) In both cases, the temporal variation of the magnetic flux {sampled by the particle's gyration} causes the increase in its perpendicular momentum.}
    \label{fig:schematic_generic}
\end{figure*}

The efficient acceleration of particles to non-thermal energies is a cornerstone of high-energy astrophysics
\citep[e.g.,][]{Kirk:2001zv, Kowal:2012rv, 2012SSRv..173..309B, Blasi:2013rva, Meszaros:2019xej, Murase:2019tjj, Matthews:2020lig}.
High-energy particles and their interactions with ambient matter, radiation, and magnetic fields give rise to multi-messenger signals, e.g., high-energy gamma rays and neutrinos, offering a unique glimpse into cosmic accelerators \citep[see, e.g.,][]{LHAASO:2023gne, LHAASO_2024, IceCube:2022der, KM3NeT:2025npi, Das:2025tfq, Das:2025vqd}. Turbulent magnetic fields and their interaction with energetic particles govern the particle transport and energization in astrophysical plasmas 
(e.g., \citealt{Fermi_1949,Kulsrud_1969,Petrosian:2004ft,Brunetti:2007zp,Keenan:2013rza, 2025ApJ...994..142H}). 
There are two basically different forms of electromagnetic acceleration mechanisms due to spatial motions of magnetic field inhomogeneities (henceforth referred to Type I) and temporal variations of magnetic field strengths (henceforth referred to Type II),  
and all known electromagnetic mechanisms can either be reduced to one of these types or are a combination of them
\citep{Ginzburg_1964}.
However, 
it is not fully understood
that which form dominates the acceleration of energetic particles in turbulence.

Particle-in-cell (PIC) simulations are advantageous in resolving the particle-turbulence interaction from first principles. Several recent studies have performed PIC simulations to investigate particle acceleration in strongly magnetized turbulence \citep[e.g.,][]{Zhdankin:2016lta, Zhdankin:2018doa, Zhdankin:2019dfz, Comisso:2018kuh, Comisso:2019frj, Comisso:2022iqy, Nattila:2020qfx, Nattila:2021qag, meringolo2023, vega2022, vega2024}.
Most earlier studies on particle acceleration focus on Type I, including 
the transit-time damping (TTD; e.g.,
\citealt{ Fermi_1949,Fermi_1954,Fisk_1976})
and pitch-angle scattering acceleration 
(e.g., \citealt{Jokipii_1966,Lemoine:2021mtv, Bresci_2022, Kempski:2023ikw, ha2024,lemoine2025}).
The TTD acceleration is characterized by the stochastic increase of particle momentum parallel to the magnetic field, and thus the stochastic decrease of particle pitch angles.
The scattering acceleration is characterized by the stochastic change of particle pitch angles and an isotropic pitch angle distribution. 

Type II acceleration remains underexplored. 
In turbulence, 
the temporal variations of magnetic field strengths are caused by turbulent compressions of magnetic fields and the Alfvénic turbulent mixing of magnetic compressions \citep{2003MNRAS.345..325C}. 
The induced 
betatron electric field
accelerates particles. 
The traditional betatron acceleration in turbulence is reversible and thus 
leads to zero net acceleration
\citep{Ginzburg_1964}.
Pitch-angle scattering is commonly invoked in Type II acceleration
to break the reversibility 
\citep{1988SvAL...14..255P, Chandran:2003xb}.
However, under the assumption of efficient scattering, Type II acceleration becomes subdominant to Type I scattering acceleration 
\citep{Cho:2005mb}.
Recently, a new Type II acceleration mechanism, the mirror acceleration, has been proposed 
\citep[][hereafter, LX23]{Lazarian:2023zui}, 
based on the earlier study by 
\citet{Cho:2005mb}. 
LX23 found that the intrinsic perpendicular superdiffusion of turbulent magnetic fields 
\citep{Lazarian:1998wd}
and thus the perpendicular superdiffusion of particles 
\citep{Lazarian:2013dba}
naturally breaks the reversibility.
Instead of invoking scattering and the associated scattering diffusion parallel to the magnetic field,
the mirroring and the associated parallel mirror diffusion  
\citep{Lazarian:2021kvd}
is found to play a key role in the mirror acceleration. 
With the perpendicular superdiffusion accompanying the parallel mirror diffusion, 
a particle can stochastically sample different compressed/expanded turbulent eddies, enabling a stochastic and efficient acceleration process. 
The mirror diffusion has been tested with numerical simulations 
(e.g., \citealt{Zhang:2023igz,Zhang:2024evq,Barreto-Mota:2024kli,Xiao:2025yrt, Lubke:2025day, 2025ApJ...994..142H}). 
The mirror acceleration results in a stochastic increase in particle momentum perpendicular to the magnetic field and in pitch angles (LX23). 
Therefore, an anisotropic pitch-angle distribution of accelerated particles, with a concentration at large pitch angles, is expected when the mirror acceleration dominates over Type I TTD and 
scattering acceleration. 

The mirror acceleration is previously studied in nonrelativistic turbulence.
In this work, 
we will extend its study to the regime of 
relativistic turbulence in strongly magnetized plasma
with the Alfven speed and the turbulent speed close to the speed of light. 
Such conditions are expected in magnetized astrophysical environments, e.g., accretion flows in active galactic nuclei (AGN), X-ray binaries, etc.
\citep[e.g.,][]{nattila2024, Groselj:2023bgy, Fiorillo:2024akm, Lemoine:2024roa}. 
We will adopt the PIC approach and perform a 3D simulation of a strongly magnetized electron-position pair plasma with driven turbulent magnetic fluctuations.  
This work will also be the first study of Type II acceleration with 3D PIC simulations. 
We will analyze the properties of accelerated particles by tracking their trajectories and statistically examining the characteristics expected for the mirror acceleration. 

We organize this article as follows. In Sec.~\ref{sec:mirror}, we briefly review the previous studies on the mirror acceleration. 
In Sec.~\ref{sec:methods},
we describe the
numerical setup and methodology of our analysis.  
We present our results in Sec.~\ref{sec:results} and discuss them in Sec.~\ref{sec:discussions}, in comparison with other Type I and Type II mechanisms. We draw our conclusions in Sec.~\ref{sec:summary}.

\section{Brief review on previous studies on the mirror acceleration\label{sec:mirror}}

\begin{figure*}
\centering
\subfigure[]{
   \includegraphics[width=0.46\textwidth]{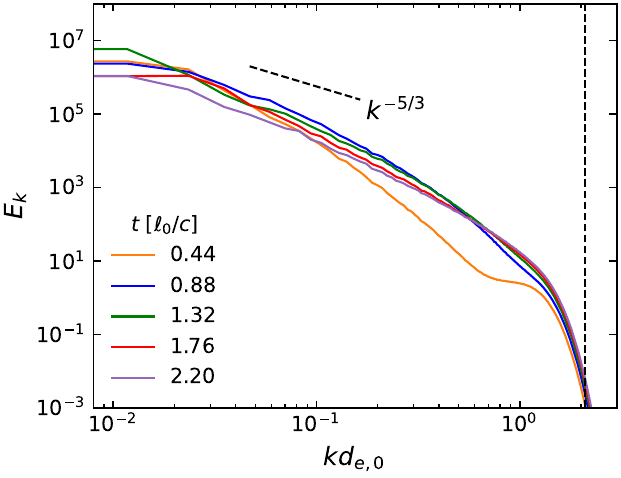}\label{fig:Ek_B}}
   \hfill
\subfigure[]{
   \includegraphics[width=0.46\textwidth]{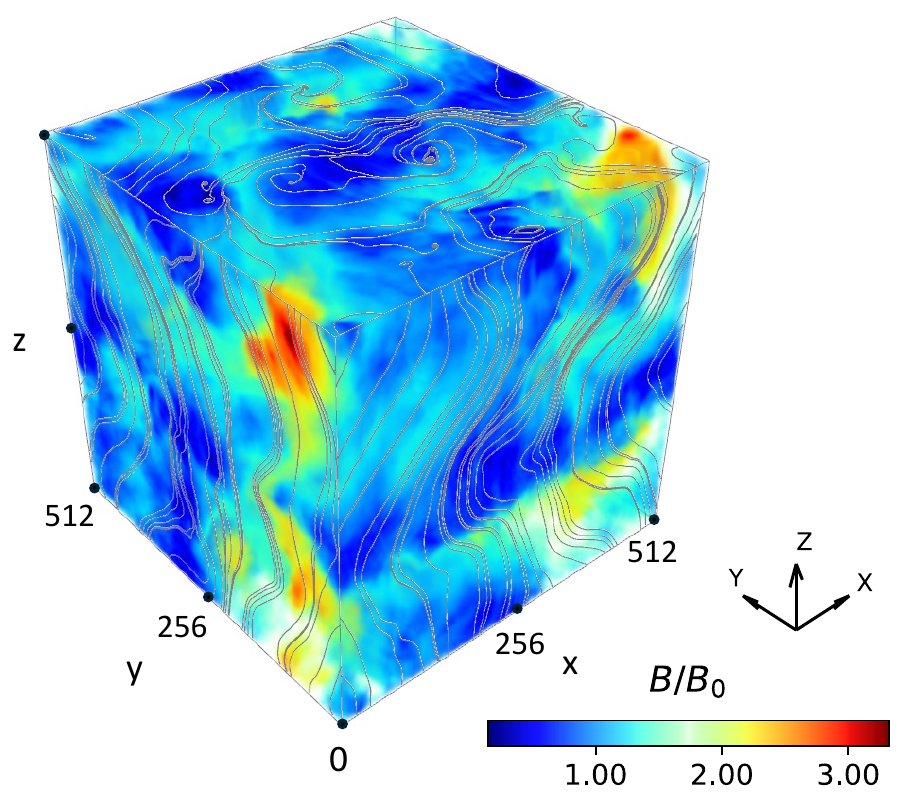}\label{fig:box_3d_b}}
\caption{(a) The magnetic energy spectrum measured at different times. The black dashed line indicates the Kolmogorov slope of turbulence. {The vertical dashed line indicates $r_{L,0}$.} (b) Distribution of magnetic field strength {at $t\approx 1.7\ell_0/c$}, normalized by $B_0$. {The grey lines indicate the magnetic field orientations.}}
\label{fig:3db}
\end{figure*}


\subsection{Mirror diffusion and mirror acceleration in nonrelativistic turbulence\label{subsec:mirror_diffusion}}

Unlike the gyro-resonant scattering, which is caused by the turbulent perturbation of magnetic field orientation, 
the non-resonant mirroring is caused by the turbulent compression of magnetic field strength. 
{The mirroring} is expected to happen in the presence of magnetic compressions, irrespective of plasma compressibility or plasma $\beta$ (ratio of gas pressure to magnetic pressure). 
{For instance, even in the limit case of incompressible plasma, 
pseudo-Alfvén modes cause compressions of magnetic fields and thus generate magnetic mirrors
\citep{Goldreich:1995sr,2003MNRAS.345..325C}.}


Turbulent compressions of magnetic fields create magnetic field gradients, i.e., magnetic mirrors, of different sizes and strengths along the turbulent energy cascade. 
While magnetic mirrors are traditionally associated with the longitudinal magnetic field gradients, 
turbulence can generate both longitudinal and transverse magnetic field gradients, and both corresponding mirroring interactions are expected. 
In the case of longitudinal mirrors,
particles with 
Larmor radius $r_L$ less than the mirror size, i.e., the scale of longitudinal magnetic field gradient, are subject to mirroring.  
The maximum value of the cosine of pitch angle, $\mu=\cos\alpha$, corresponding to the smallest pitch angle that allows mirror reflection, is given by $\mu < \sqrt{\delta B / (B_0 + \delta B)}$, where $\alpha$ is the angle between the particle velocity and the magnetic field, and $B_0$ and $B_0 + \delta B$ are the magnetic field strengths in the weak and strong field regions, respectively. A particle satisfying the above conditions 
has its parallel motion with respect to the local magnetic field 
reversed at the mirror reflection. 
The mirroring naturally overcomes the theoretical $90^\circ$ problem of the quasi-linear theory (QLT) for scattering\footnote{The scattering vanishes when the particle pitch angle is close to $90^\circ$, which is known as the $90^\circ$ problem of the QLT \citep{Jokipii_1966}.}
\citep{Cesarksky_1973}.

In the case of transverse mirrors,
a particle from a weaker field is reflected after a half-gyration around the stronger field and directed back to the weaker field 
(Fig.~\ref{fig:schematic_generic}). 
The mirror size, i.e., the scale of transverse magnetic field gradient, is comparable to $r_L$.

The mirror diffusion in 
\citet{Lazarian:2021kvd}
is discussed for the case of longitudinal mirrors.
Unlike the trapping of particles between two magnetic mirrors 
in linear MHD waves \citep{Cesarksky_1973}, in nonlinear MHD turbulence with perpendicular superdiffusion of turbulent magnetic fields \citep{Lazarian:1998wd}, 
particles also undergo perpendicular superdiffusion 
\citep{Xu:2013ppa,Lazarian:2013dba,Hu_2022,Zhang:2023igz}
and 
stochastically encounter different mirrors as they travel along the magnetic field.
Therefore, they undergo mirror diffusion in the direction parallel to the local magnetic field, as proposed by \cite{Lazarian:2021kvd} and 
numerically tested by e.g., 
\cite{Zhang:2023igz,Barreto-Mota:2024kli,Xiao:2025yrt, Lubke:2025day, 2025ApJ...994..142H}.

%
%
The mirror acceleration, as formulated in LX23 based on the mirror diffusion, is in the nonrelativistic turbulence regime. 
In nonrelativistic turbulence, the mirror acceleration is attributed to the large-scale turbulent compressions
\citep{Cho:2005mb}, 
which correspond to larger turbulent energies compared to the $r_L$-scale eddies. 
To sample large-scale turbulent compressions, the parallel mirror diffusion of particles via interactions with 
longitudinal mirrors is important, 
as the displacement of parallel diffusion and that of perpendicular superdiffusion are related by the anisotropic turbulent scaling 
\citep{Goldreich:1995sr,Lazarian:1998wd,Lazarian:2023sh}. 
The mirror diffusion solves the theoretical difficulties faced by the mirror trapping and scattering diffusion adopted in earlier studies of 
Type II acceleration.
A particle trapped in the same eddy would undergo oscillatory compression and expansion of magnetic fields, and thus, the crucial stochasticity for stochastic acceleration would be lost, i.e., a reversible acceleration with zero net energy gain. 
Trapping would also inhibit the particle from sampling large-scale turbulent compressions. 
If scattering and pitch-angle isotropization were efficient, Type II acceleration would become subdominant to Type I scattering acceleration
\citep{Cho:2005mb}. 
Moreover, 
particles would simultaneously sample the magnetic compression in one direction and expansion in another direction in incompressible plasma,
with zero net energy gain
\citep{Cho:2005mb}.
The mirror diffusion does not cause pitch-angle isotropization, and 
a net energy gain is expected, 
irrespective of the plasma compressibility. 


The mirror acceleration 
leads to a stochastic increase in the particle momentum perpendicular to the magnetic field $p_\perp$ and thus in $r_L$. It causes the stochastic decrease of $\mu$ and thus is self-sustained with the longitudinal mirroring
condition always satisfied. 

\subsection{Mirror acceleration in relativistic turbulence}
\label{ssec:marel}

In relativistic turbulence, 
the stochastic acceleration becomes much more efficient than its non-relativistic counterpart 
\citep{Dermer_2009}. 
Compared to the nonrelativistic turbulence, 
the $r_L$-scale turbulent compression can cause a rapid 
magnetic flux change and a significant violation of the first adiabatic invariant ($J_1 \propto p_\perp^2/B$) 
for a particle gyrating around a transverse mirror. 
Therefore, the interaction with $r_L$-scale transverse mirrors becomes important in the mirror acceleration in relativistic turbulence. 
As illustrated in Fig.~\ref{fig:schematic_generic},
irrespective of the reflection of parallel momentum by a longitudinal mirror, for a particle interacting with a transverse mirror, the temporal variation of the magnetic flux {sampled by its gyration} can cause its acceleration by the induced betatron electric field. 
Given the relativistic turbulent compression, 
each transverse mirror interaction can cause a significant energy gain and thus a non-closed gyro-orbit. 
For mirror acceleration in relativistic turbulence, 
the reversibility is broken due to the turbulent perpendicular superdiffusion, nonconserved $J_1$,
and distorted gyro-orbits of particles. 
The mirror acceleration results in a stochastic increase in $p_\perp$ and $r_L$. 
With the stochastic increase in $r_L$ and pitch angles, the transverse mirror interactions are more favored than the longitudinal ones for the mirror acceleration in relativistic turbulence.


\section{Numerical simulation} \label{sec:methods}

We perform a 3D PIC simulation of strongly magnetized turbulence using the open-source plasma simulation framework RUNKO \citep{2022A&A...664A..68N}. It solves the relativistic Boltzmann-Vlasov equations for particles in six-dimensional phase space. The electron-positron pair plasma is initially warm with a temperature parameter $\Theta=k_BT/m_ec^2=0.3$ and uniform with a total number density $n_e$ of electrons and positrons, where $k_B$ is the Boltzmann constant, $T$ is the temperature, $m_e$ is the electron rest mass, and $c$ is the speed of light. Here we focus on the relativistic scenario and set the plasma magnetization parameter $\sigma_0=B_0^2/4\pi n_e m_ec^2 =10$, where $B_0$ is the mean magnetic field. The spatial resolution is $512^3$ cells. 
{The initial plasma skin depth $d_{e,0} = c/\omega_p$ is resolved with 3 cells}, where $\omega_p=\sqrt{4\pi n_ee^2/m_e}$ is the plasma frequency, and $e$ is the electron charge. 
{The initial thermal particles have the averaged Larmor radius $r_{L,0} \approx 1.4$ cells,
which becomes better resolved as the particles are energized.}

We adopt the same turbulence driving method as detailed in \cite{nattila2024}. 
The plasma is initialized with a uniform magnetic field $\bm{B_0}$, and the turbulent magnetic fluctuation is continuously driven such that the total magnetic field is $\bm{B}=B_0\hat{z}+\delta \bm{B}$, where $\delta\bm{B}$ is the driven magnetic fluctuation in the direction perpendicular to the mean field, with the strength $\delta B\sim B_0$. The turbulence driving scale $\ell_0$ is set to be equal to the box size to maximize the turbulence inertial range. 


Fig.~\ref{fig:Ek_B} presents the time-evolving magnetic energy spectrum. We find that over $t\approx2\ell_0/c$, the energy spectrum gradually approaches the Kolmogorov spectrum ($\sim k^{-5/3}$)
expected for a turbulent cascade. 
At later times, some spectral flattening is seen  
near wavenumbers $k=1-2$. 
This low-k spread might be related to the anisotropy in driving, and further work is needed for its explanation.
We stop the run before the magnetic energy spectrum significantly steepens due to 
the dissipation of magnetic fluctuation energy into the thermal and kinetic energies of particles. 
{As an example, Fig.~\ref{fig:box_3d_b} shows the magnetic field strength distribution at a time 
$t\approx 1.7\ell_0/c$.
The overlaid lines indicate the magnetic field orientations.}
Turbulence induces fluctuations in magnetic field strength, naturally giving rise to magnetic mirrors.

\begin{figure}
    \centering
    \includegraphics[width=0.48\textwidth]{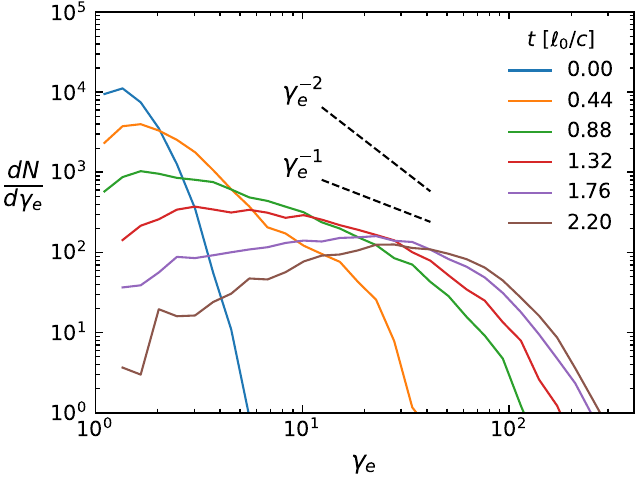}
    \caption{Energy spectrum of particles tracked through the simulation. A non-thermal component gradually develops at later time. The black dashed lines show two reference power-law scalings.}
    \label{fig:dndg}
\end{figure}

\begin{figure*}
\centering
\subfigure[]{
   \includegraphics[width=0.48\textwidth]{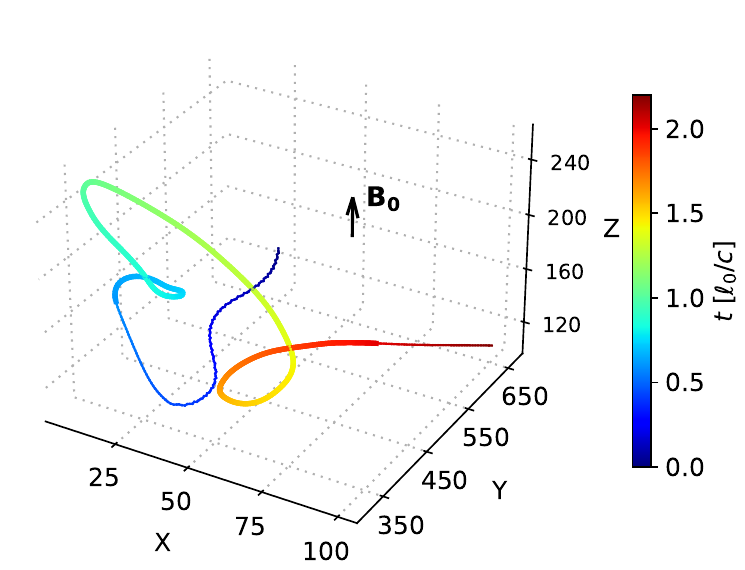}\label{fig:trajectory_t}}
\subfigure[]{
   \includegraphics[width=0.48\textwidth]{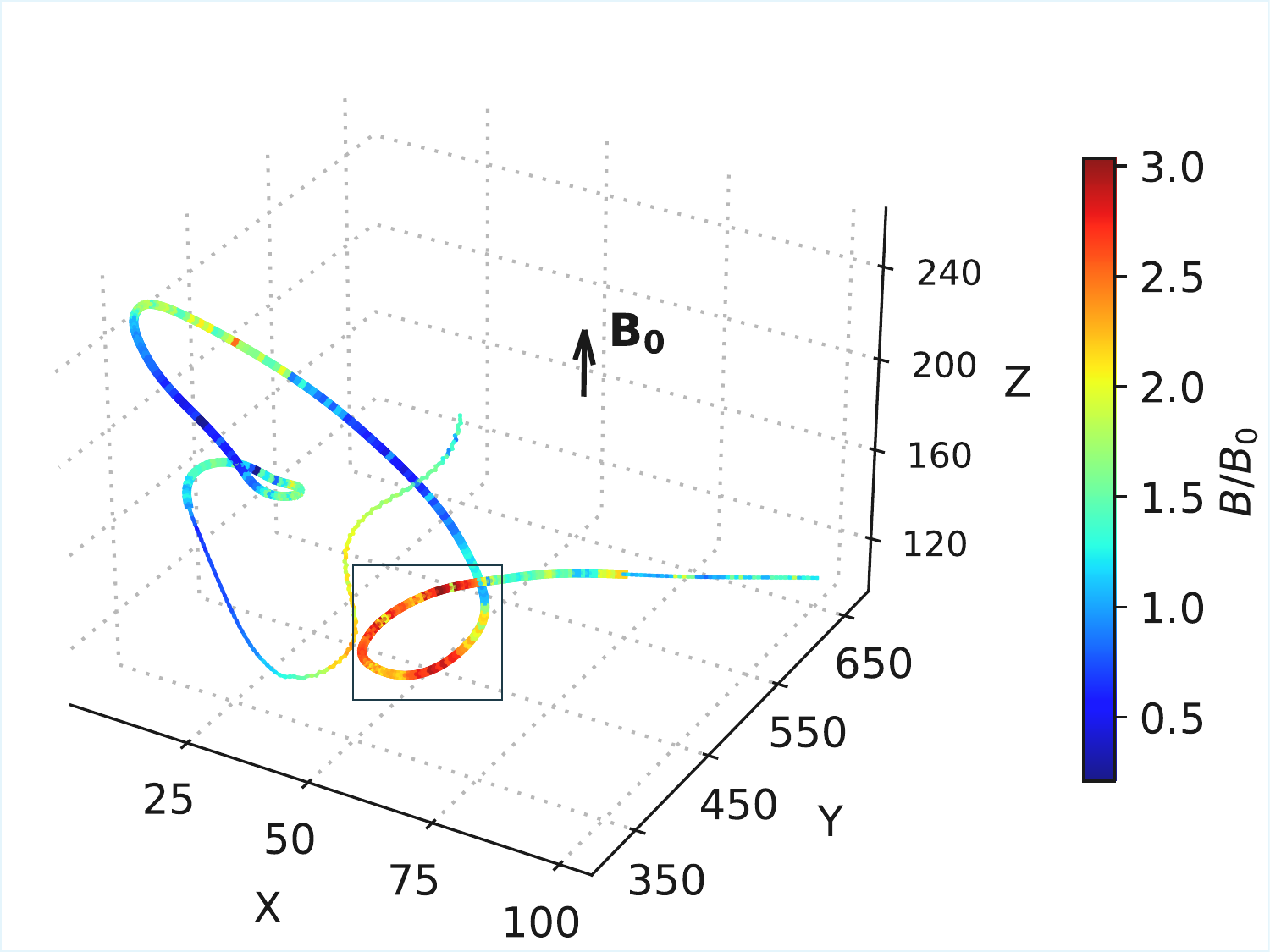}\label{fig:trajectory_b}}
\caption{(a) Trajectory of a particle tracked in the simulation, color coded by time. The thickened segment corresponds to the time interval $t=0.6-2.0\ell_0/c$, with 
multiple mirror interactions. 
The mean magnetic field direction (z direction) is indicated. (b) Same as (a), but color-coded by $B/B_0$. The segment marked by the black square exemplifies a mirror interaction, shown in the zoomed-in view in Fig.  \ref{fig:mirror_zoom}.}. 
\label{fig:trajectory}
\end{figure*}

The total number of particles in the simulation is $\approx 2\times 10^{9}$. We track an ensemble of $10^4$ particles randomly selected to study their acceleration.
This sample size is sufficient for us to achieve statistically significant and convergent results.
We measure quantities including the particle Lorentz factor ($\gamma_e$), parallel ($p_\parallel$) and perpendicular ($p_\perp$) components of momentum, 
$r_L (= p_\perp c/e B$), $\mu$, and the local electric and magnetic fields. 
We also perform a Lorentz transformation 
to boost the quantities to the frame moving with the $\bm{E}\times\bm{B}$ drift velocity 
(see Appendix \ref{app:drift}). This allows us to 
remove the energy oscillations associated with particle gyrations seen in the laboratory frame 
\citep{ Comisso:2019frj, Wong:2019dog, Bresci_2022}. Through this paper, primed quantities refer to those measured in the drift velocity frames, while unprimed quantities refer to those in the laboratory frame. 

\section{Results} \label{sec:results}

\subsection{Temporal evolution of particle energy spectrum}

 
We measure the energy spectrum of the sampled particles tracked through our simulation at different times, as shown in Fig.~\ref{fig:dndg}.
The overall distribution broadens with time by more than one order of magnitude toward the end of the simulation. Starting from the initial thermal distribution, a significant non-thermal power-law tail starts to develop at $t\approx 0.4$~$\ell_0/c$ (orange solid line). 
The non-thermal tail gradually approaches $dN/d\gamma_e \propto \gamma_e^{-1}$, as generally expected for stochastic acceleration in the absence of cooling and escaping of particles \citep[e.g.,][]{Melrose_1969, Longair_2011, Xu:2017ypi}. 
Toward the end of the simulation, as the extension of the non-thermal tail is limited by the box size, and meanwhile, the overall thermal energy
increases with the thermal bump shifting to progressively higher energies, the non-thermal tail shortens,
steepens, and eventually is truncated. We note that in realistic {cosmic} accelerators, with the accelerator size much larger than our numerical box size,
a much more extended non-thermal power-law spectrum is expected.

\subsection{Particle trajectory and acceleration properties\label{subsec:trajectory}}

\begin{figure*}
    \centering
    \includegraphics[width=0.98\textwidth]{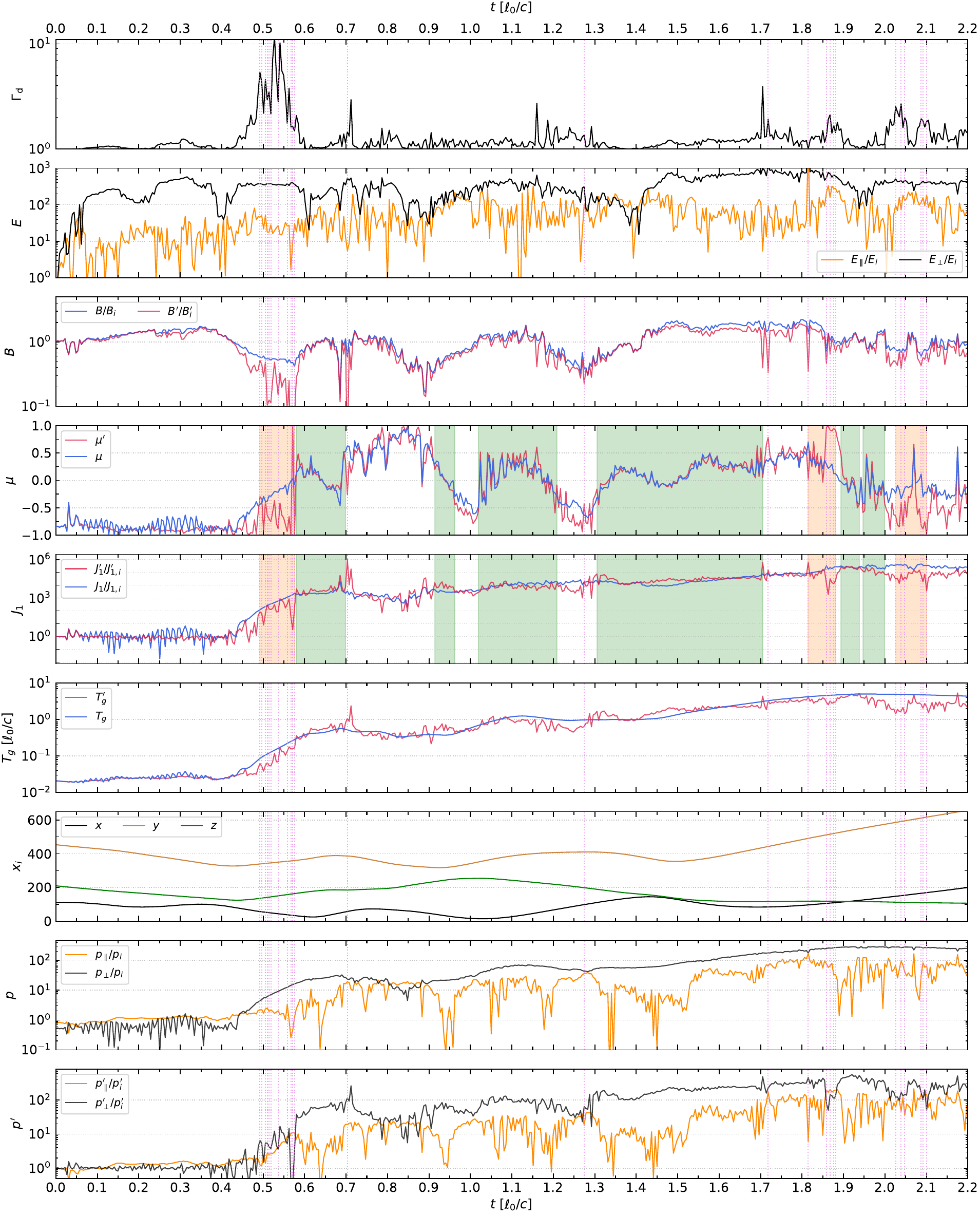}
    \caption{The quantities measured along the particle trajectory shown in Fig. \ref{fig:trajectory} in the laboratory (unprimed) and {drift velocity} (primed) frames. 
    The subscript $i$ indicates the initial values measured for the trajectory. 
    The vertical dotted magenta lines correspond to $E>B$. 
    {The green and orange-shaded regions in $\mu$ and $J_1$ panels correspond to the identified mirror and reconnection intervals, respectively.}
    }
    \label{fig:drift_frame}
\end{figure*}

As an example, Fig.~\ref{fig:trajectory_t} displays the trajectory of a particle tracked in the simulation. The physical quantities, namely the electric field $E$, the magnetic field $B$, $\mu$, $J_1$, 
{the gyroperiod $T_g = 2\pi r_L/c$,} $p_\parallel$, $p_\perp$, and the trajectory coordinates $x$, $y$, and $z$ are presented in Fig.~\ref{fig:drift_frame}. 
We note that  
{the gyroperiodic oscillations in $p_\perp$ due the 
effect of $\bm{E}\times\bm{B}$ drift is prominent at $t<0.5 \ell_0/c$
when the particle energy is low}.
At early times, the gyrations with $r_L$ of a few cells are hardly discernible in Fig.~\ref{fig:trajectory_t}. At a later time, $r_L$ significantly increases. 
The particle undergoes several reversals in directions  perpendicular and/or parallel to the magnetic field, 
with distorted gyrations. 
{Toward the end of the simulation, 
{$r_L$ becomes comparable to the box size (Fig. \ref{fig:drift_frame}), 
and the particle travels approximately perpendicular to the box-averaged magnetic field ($z$-direction)}. }

In Fig.~\ref{fig:drift_frame}, we see small variations in the drift Lorentz factor $\Gamma_d$, interspersed with occasional large variations. By further comparing the magnetic and electric field strengths, we find that most of the spikes in $\Gamma_d$ occur at $E>B$. The condition $E>B$ indicates the local weakening of reconnecting magnetic fields. 
The corresponding acceleration by reconnection causes an increase in 
$p_\perp$ around $t\approx0.5\ell_0/c$. 

In earlier studies on both turbulent acceleration and reconnection acceleration, the acceleration by non-ideal electric fields is identified, 
but it is not the dominant acceleration mechanism for particle energization \citep[e.g.,][]{Guo:2019acp, Comisso:2019frj}. 
{For reconnection emerging in turbulence, turbulent flow can induce motional electric field and contribute to $E_\perp$($>E_\parallel$) 
observed in local reconnection regions}, where $E_\perp$ and $E_\parallel$ are the components of $E$ 
perpendicular and parallel to the local magnetic field. 
{The dominant increase in $p_\perp$ suggests that the reconnection acceleration is 
primarily driven by the motional electric field.}

Later in the range $t\approx 0.6-2.0\ell_0/c$,  
devoid of magnetic reconnection, 
we see that the particle repeatedly reaches large $\alpha$, i.e., small $|\mu|$.
The corresponding reversals of particles in space 
can be seen from the $x$, $y$, and $z$ components of the particle position.
Meanwhile, the particle reversals are accompanied by significant acceleration, with further increase of $r_L$, until $r_L$ reaches a scale comparable to the box size. 
With $p_\perp>p_\parallel $ mostly seen, it is clear that the acceleration preferentially takes place in the direction perpendicular to the magnetic field. 
Consequently, the particle that initially moves along the magnetic field with a large $|\mu|$, moves approximately perpendicular to the magnetic field with $\mu$ close to 0 after acceleration, as also seen in Fig.~\ref{fig:trajectory}.

We note that at $t>1.7\ell_0/c$, the particle again encounters reconnection regions. 
In the absence of cooling, the thickness of current layers determined by the gyroradius of thermal particles \citep[e.g.,][]{Guo:2016yfq} increases with time (see 
Appendix~\ref{app:jdist}). Toward the end of the simulation, the regions of reconnection become more volume-filling, and thus the accelerated particle more frequently traverse the reconnection regions.
{However, 
a particle encounter with a reconnection region does not necessarily correspond to reconnection acceleration,
when $r_L$ far exceeds the size of reconnection region}.
In realistic situations with well-separated scales, the chance of a highly energetic particle encountering  microscopic reconnection layers in turbulence is expected to be low.

\subsection{Identifying mirror interactions}\label{ssec:identi}
\begin{figure*}
\centering
\subfigure[]{
   \includegraphics[width=0.48\textwidth]{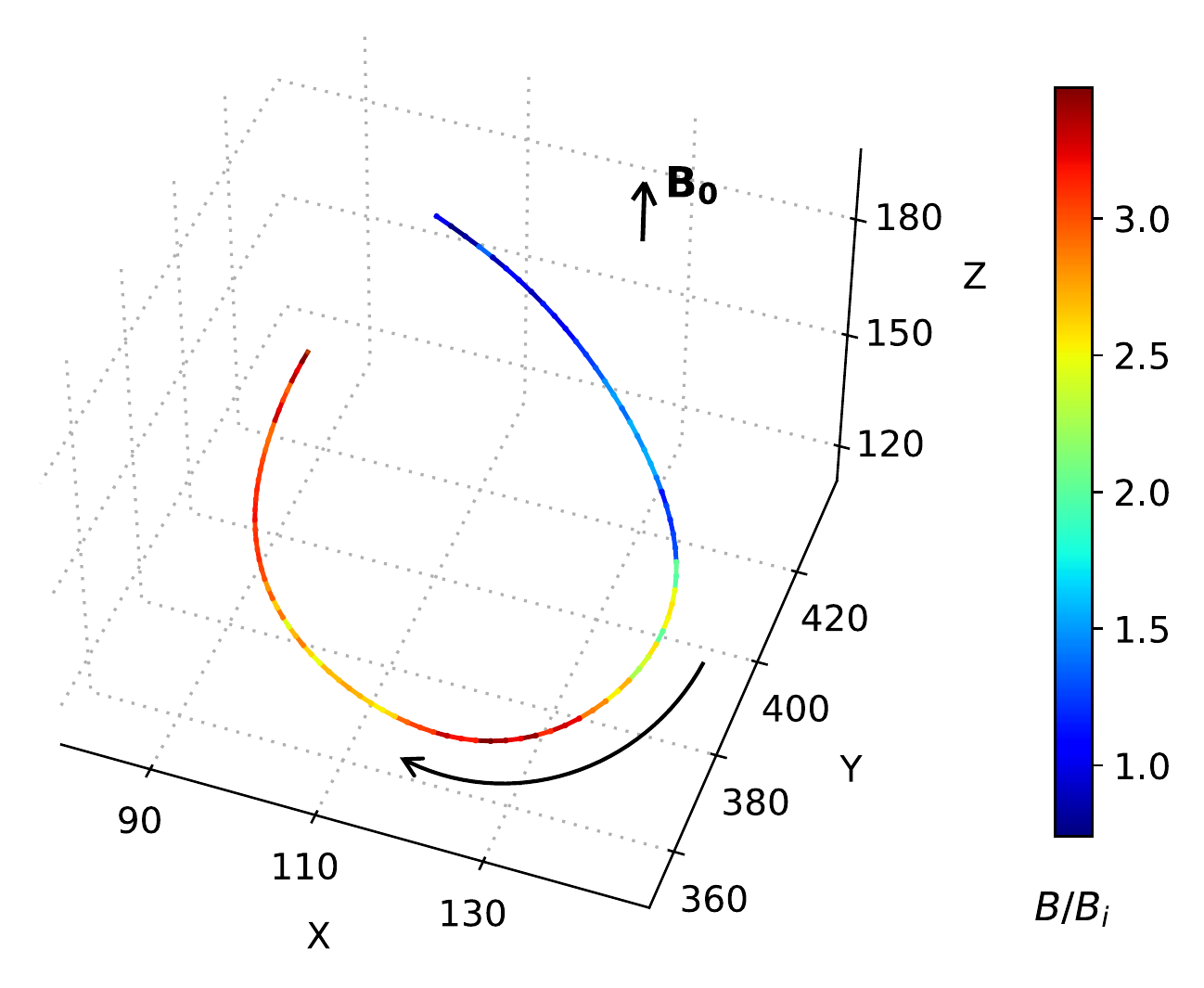}\label{fig:bcolor_seg}}
   \hfill
\subfigure[]{
   \includegraphics[width=0.48\textwidth]{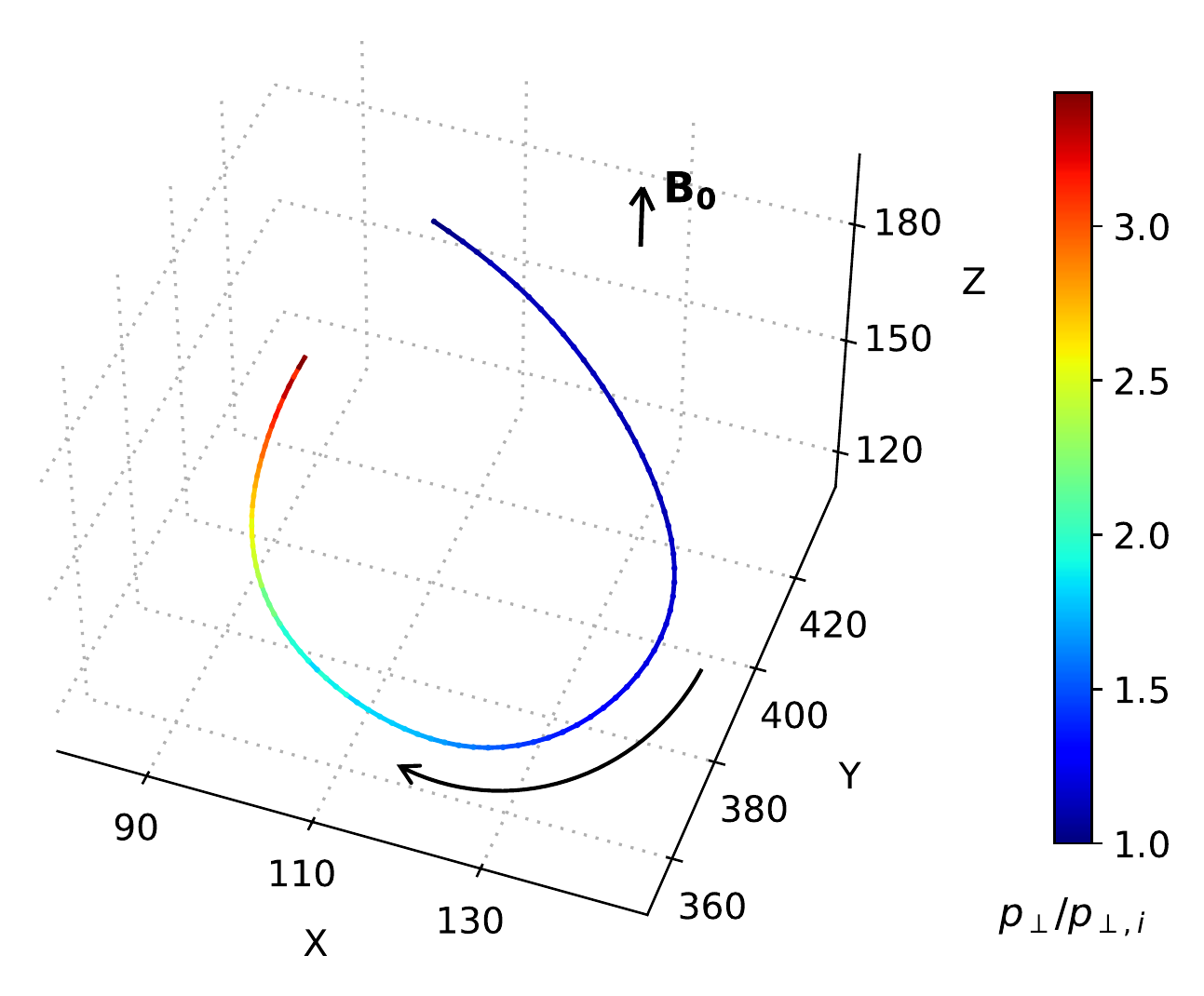}\label{fig:pcolor_seg}}
\caption{Zoomed-in view of the particle trajectory in the time interval $t\approx 1.3-1.7\ell_0/c$ marked by the black square in Fig.~\ref{fig:trajectory_b}. The curved arrow indicates the particle's moving direction. (a) and (b) are color-coded by $B$ and $p_\perp$, respectively, which are normalized by their initial values of this trajectory segment.}
\label{fig:mirror_zoom}
\end{figure*}

To more closely examine the acceleration mechanism
during the time 
devoid of magnetic reconnection, 
we focus on the 
interval $t\approx 1.3-1.7\ell_0/c$
with small $|\mu|$
as an example, 
as marked in Fig.~\ref{fig:trajectory_b}. Its zoomed-in view is presented in Fig.~\ref{fig:mirror_zoom}. 
With $\mu$ close to 0, the particle gyrates around the magnetic field while experiencing the strengthening of the field. The spatial variation of $B$, i.e., the transverse magnetic field gradient, 
{reverses the particle's moving direction perpendicular to the magnetic field.} 
Meanwhile, the temporal variation of $B$ induces an electric field that accelerates the particle, causing a significant increase in $p_\perp$. 
With a systematic variation of $J_1$ {within $T_g$} during the mirroring interaction as shown in Fig. \ref{fig:drift_frame},
{the acceleration mechanism is different from the scattering acceleration with a stochastic variation of $J_1$.}  
The increase in $p_\perp$ results from 
the temporal variation of the magnetic flux 
{sampled via the particle's gyration.}
This is the mirror acceleration mechanism described in Sec.~\ref{ssec:marel} (see Fig.~\ref{fig:schematic_generic}).

In the presence of relativistic turbulence, significant energy gain can happen during one mirror interaction within one gyro-orbit. 
It follows that a mirror-accelerated particle is characterized by incomplete and distorted gyrations (see Fig. \ref{fig:trajectory}).
We note that mirror acceleration is a stochastic acceleration process. When the local magnetic field is expanding, and $B$ is weakening over time, a mirror interaction can lead to a decrease in $p_\perp$.

The particles with small $|\mu|$
can most effectively sample the temporal change of magnetic flux via their gyrations. They 
undergo the mirror acceleration that further causes the stochastic increase in $p_\perp$ and decrease in $|\mu|$. 
The time range with repeated mirror interactions is indicated by 
the thickened trajectory in Fig.~\ref{fig:trajectory}.

To systematically identify the mirror interactions 
for each sampled particle, 
we select the time range $t>0.5\ell_0/c$ when the non-thermal population is developed. 
We further exclude reconnection acceleration by selecting the time intervals for analysis
with durations
$\Delta t>0.04\,\ell_0/c$ (i.e., $>10$ consecutive simulation time steps)
and continuously $E<B$.
Within these time intervals, we then identify a mirror interaction when $|\mu^\prime|<0.4$ is continuously observed over a duration of $\Delta t\geq 0.04\ell_0/c$. 
We adopt the drift velocity frame to avoid any drift effect on the identification. 
We focus on transverse mirrors and therefore do not require $\mu^\prime$ to cross $0$, 
as the mirror acceleration can occur regardless of the reflection 
by a longitudinal mirror (see Fig. \ref{fig:schematic_generic}).
To avoid counting transient fluctuations and spurious spikes 
of $\mu^\prime$ across $|\mu^\prime| = 0.4$, 
we impose $\Delta t\geq 0.04\ell_0/c$, and  
we also smooth the original $\mu^\prime$ data by 
applying a 5-point moving average technique \citep{oppenheim99}
prior to the identification of mirror interactions.

As an example, the mirror interactions identified with the above method for the particle shown in Fig.~\ref{fig:drift_frame}
are indicated by the {green} shades in the panels of $\mu$ and $J_1$. 
$J_1$ measured during each mirror interaction shows a systematic variation. 
It suggests that the mirror interaction is accompanied by the temporal change of magnetic flux sampled by the particle.

\subsection{Statistical analysis on   
mirror acceleration}
\label{subsec:statistics}

\begin{figure}
\centering
\includegraphics[width=0.48\textwidth]{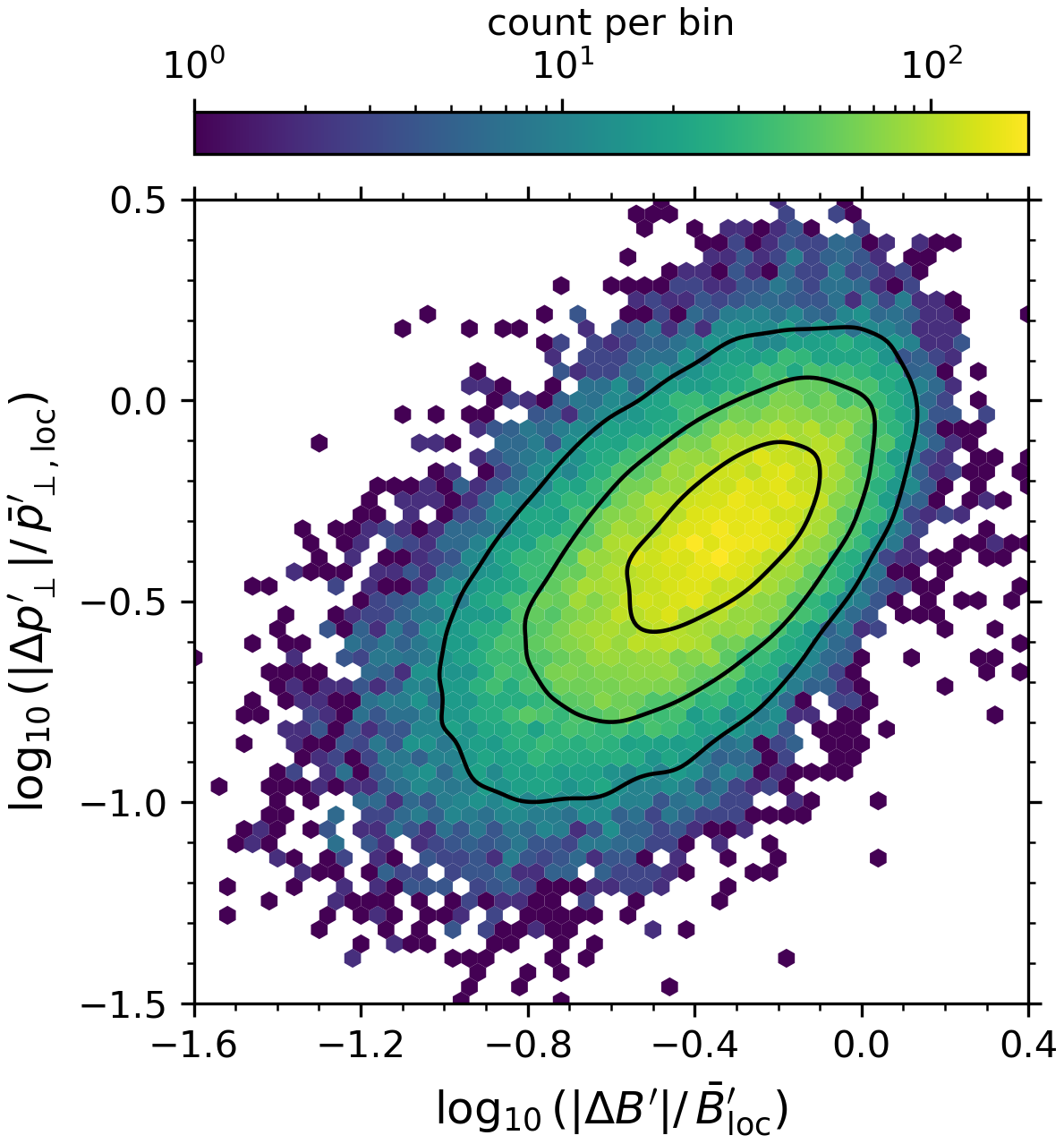}
\label{Fig:fig7a}
\caption{2D distribution plot of $|\Delta p^\prime_\perp |/\,\bar{p}^\prime_{\perp,\text{loc}}$ and $|\Delta{B}^\prime|/\,\bar{B}^\prime_\text{loc}$ for the $39519$ identified mirror interactions  
from the $10^4$ sampled particles.
The three contours enclose 0.3, 0.7, and 0.9 of the total counts around the distribution peak.
{The hexbin spacings} along the $x$- and $y$-axes on a logarithmic scale 
are $0.04$ and $0.07$, respectively.}
\end{figure}

From the $10^4$ sampled particles, 
we identify $39519$ mirror interactions in total with the method described in Section \ref{ssec:identi}.
As the mirror acceleration occurs due to the temporal variations of magnetic field strengths, 
the change in $p_\perp$ depends on the local magnetic compression/expansion
\citep[][LX23]{Cho:2005mb}.
We measure the fractional change in $p_\perp$ as 
$|\Delta p^\prime_\perp|/\bar{p}^\prime_{\perp,\text{loc}}$
and examine its correlation with the relative change of the local magnetic field strength 
$|\Delta B^\prime|/\bar{B}^\prime_\text{loc}$. 
Here $|\Delta p^\prime_\perp| = p^\prime_{\perp,\text{max}}-p^\prime_{\perp,\text{min}}$,
where $p^\prime_{\perp,\text{max}}$
and $p^\prime_{\perp,\text{min}}$
are the maximal and minimal $p^\prime_\perp$
during each identified mirror interaction, and 
$\bar{p}^\prime_{\perp,\text{loc}}$ is the local averaged $p^\prime_\perp$ during the mirror interaction. 
We note that for a stochastic acceleration, $\Delta p^\prime_\perp$ can be either positive or negative.
Moreover, 
$|\Delta B^\prime| = B^\prime_\text{max}-B^\prime_\text{min}$, where $B^\prime_\text{max}$ and $B^\prime_\text{min}$ are the maximal and minimal magnetic field strength during the corresponding mirror interaction, and 
$\bar{B}^\prime_\text{loc}$ is the local averaged magnetic field strength during the mirror interaction. 
We adopt the drift velocity frame to avoid the effect of drift on gyrations 
and thus isolate the sampling of magnetic flux change through gyrations alone. 
We also smooth the original $p^\prime_\perp$ and $B^\prime$ data 
by applying the 5-point moving average technique to avoid spurious spikes when measuring their changes.

In Fig.~\ref{Fig:fig7a}, we present the resulting distribution plot of 
$|\Delta p^\prime_\perp |/\,\bar{p}^\prime_{\perp,\text{loc}}$ and $|\Delta{B}^\prime|/\,\bar{B}^\prime_\text{loc}$ for all the identified mirror interactions and find  
a Pearson coefficient of $r\approx 0.5$. 
Their positive correlation is expected for the mirror acceleration. 
Three contour levels indicate regions near the distribution peak, containing 0.3, 0.7, and 0.9 of the total counts of mirror interactions. 

{To examine the relative contribution of mirror and reconnection acceleration in turbulence to  energization of the highest-energy particles, 
we select $2875$ particles 
from the $10^4$ sampled particles
that reach $\gamma_e\geq 100$ within the 
simulation time. 
They represent the most efficiently accelerated particle population.
To identify time intervals of reconnection, 
we first identify individual $E>B$ occurrences. 
The neighboring occurrences separated by less than 
$0.04\,\ell_0/c$ 
are combined into a single interval.
We then select only intervals with durations 
$\Delta t\geq 0.04,\ell_0/c$ as reconnection intervals, 
to avoid counting transient $E>B$ fluctuations. 
As an example, 
the orange shades in 
Fig.~\ref{fig:drift_frame} 
indicate the selected reconnection intervals.
As discussed above, 
the selected reconnection intervals do not necessarily correspond to reconnection acceleration.
For high-energy particles with $r_L$ much larger than the size of the reconnection region,
insignificant energy change during the reconnection interval is expected. 
Low-energy particles with $r_L$ smaller than the smallest turbulent eddy size 
can undergo an impulsive kick by $E_\perp$ in a reconnection region. 
Accompanying the boost in $p_\perp$, a fast rise in $J_1$ is seen in 
Fig.~\ref{fig:drift_frame}.  
Different from the mirror acceleration, for which the energy gain depends on the particle's current energy,
the reconnection acceleration does not depend on the initial particle energy and is usually treated as a pre-acceleration or injection process
\citep{Comisso:2018kuh, Guo:2025zso}
for the subsequent acceleration with $r_L$ in the inertial range of turbulence.
As the reconnection in turbulence is regulated by turbulence dynamics 
\citep{Lazarian:1998wd},
despite the different acceleration mechanisms, 
we expect that the reconnection acceleration and mirror acceleration are intrinsically coupled via the turbulent energy cascade, 
which will be investigated in our future work.} 

\begin{figure}
    \includegraphics[width=0.46\textwidth]{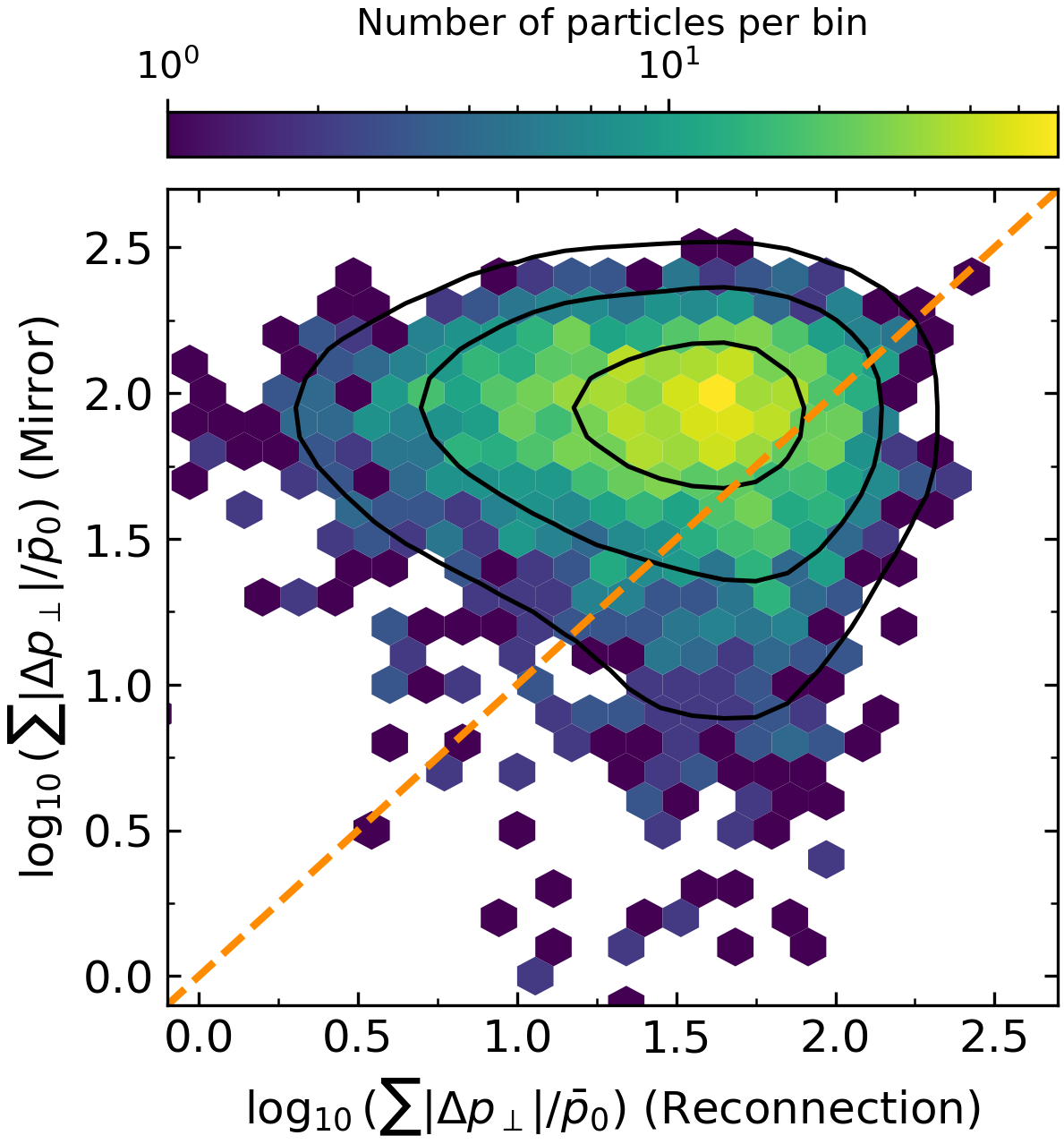}
    \caption{{2D distribution plot of  
    $\sum|\Delta p_{\perp}|/\bar{p}_{0}$
    over the mirror and reconnection intervals of the most energized particles ($\approx 2500$), where $\bar{p}_{0}$ is their averaged initial momentum.
    The three contours enclose 0.3, 0.7, and 0.9 of the total number of particles around the distribution peak. 
    The hexbin spacings along the $x$- and $y$-axes on a logarithmic scale are 0.11 and 0.20, respectively. The diagonal dashed line marks equal $\sum|\Delta p_\perp|$ from mirroring and 
    reconnection.}}
    \label{fig:mirr_recon}
\end{figure}

{As a crude estimate, 
for each of the selected particles with both mirror and reconnection intervals 
($\approx 2500$), 
we compare the sum of absolute changes in $p_\perp$,  
$\sum|\Delta p_{\perp}|$,
over all the mirror intervals with that over all the reconnection intervals. 
For the purpose of comparison, we adopt the lab frame for $\Delta p_\perp = p_{\perp, \text{max}} - p_{\perp,\text{min}}$ measurements, 
where $p_{\perp, \text{max}}$ and 
$p_{\perp, \text{min}}$ are the maximal and minimal $p_\perp$ during each identified interval.
In the lab frame, we can also avoid the spurious variations in  
$p^\prime_\perp$ 
caused by changes in the drift-frame transformation 
(Eq.~\eqref{eqn:frame})
when $E>B$ and $E<B$ occurrences are interspersed within a reconnection interval.
As shown in Fig.~\ref{fig:mirr_recon}, 
the distribution indicates a larger contribution from mirror acceleration in the particle energization, 
with $\approx 76\%$ of the 
particles on the mirror-dominated side.} 

\begin{figure*}
    \centering
    \subfigure[Mirror intervals]{
    \includegraphics[width=0.46\textwidth]{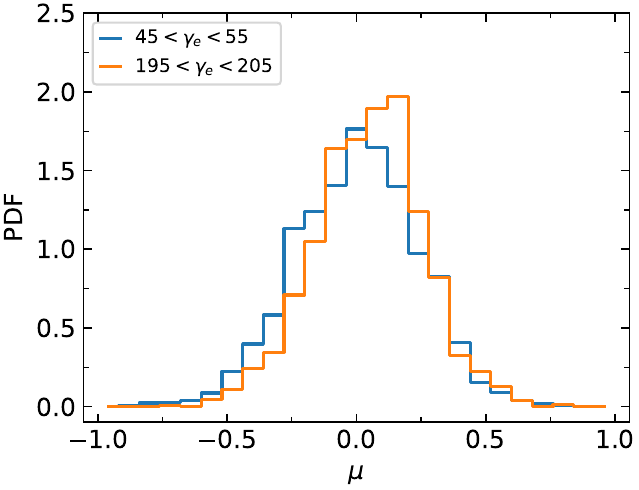}
    \label{fig:mu_hist_mirror}
    }%
    \hfill
    \subfigure[All time steps]{\includegraphics[width=0.46\textwidth]{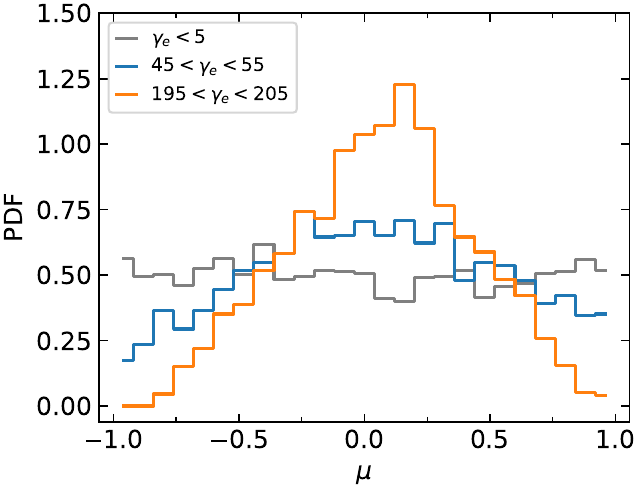}
    \label{fig:mu_hist_all}
    }
    \caption{ 
    Probability density functions (PDFs) of $\mu$ for different ranges of $\gamma_e$, 
    {based on (a) the time steps within mirror intervals and (b) all time steps. 
    Each PDF uses the same number ($\approx 2100$) of randomly sampled data points.}} 
\end{figure*}

The stochastic increase in $p_\perp$
caused by the mirror acceleration indicates a stochastic decrease in $|\mu|$ for the accelerated particles. 
Therefore, for energization dominated by mirror acceleration, the accelerated particle distribution is expected to become increasingly concentrated toward $\mu\approx0$ at higher energies. 
{In Fig.~\ref{fig:mu_hist_mirror}, 
based on the $(\mu,\gamma_e)$ values from the time steps within the identified mirror intervals of all sampled particles ($10^4$), 
we show the pitch angle distributions for different ranges of $\gamma_e$.
We see that the mirror acceleration is characterized by an anisotropic particle distribution concentrated at small $|\mu|$. 
As a comparison, 
Fig.~\ref{fig:mu_hist_all} shows the pitch angle distributions 
from all time steps without selection.}
As $\gamma_e$ increases, the distribution becomes increasingly anisotropic, shifting toward a 
concentration near $\mu\approx 0$
at $\gamma_e \approx 200$.
This trend indicates the dominance of mirror acceleration in the energization of high-energy particles.

By preferentially accelerating particles in the direction perpendicular to the magnetic field, 
the mirror acceleration 
enables repeated interactions with transverse mirrors 
within a limited-size accelerator, further enhancing the mirror acceleration. 

%


\section{Discussion} \label{sec:discussions}

\subsection{Comparison with other candidates of turbulent acceleration mechanisms}

{ (1) Scattering acceleration (Type I): }
Via the gyroresonant scattering, 
if the scattering is sufficiently efficient, 
particles can gain energy from the wave motions in turbulence
\citep{Jokipii_1966,Schlickeiser_2002}.
Recent studies on 
scattering in MHD turbulence, especially in the case with a negligible mean field, 
find that scattering by 
``sharp bends of magnetic field lines"
\citep{Lemoine:2023sxw}
can result in large-pitch-angle scattering \citep[e.g.,][]{Kempski:2023ikw,Zhang:2024evq}
and locally efficient scattering acceleration \citep{Bresci_2022,lemoine2025}.
However, the pitch angle isotropization expected for efficient scattering acceleration 
cannot explain the anisotropic pitch angle distribution of energetic particles 
seen in this work and previously in e.g., 
\cite{Comisso:2019frj,Comisso:2021isw, Comisso:2024iyx,Pezzi:2021bvc}.

(2) TTD and gradient-B drift acceleration (Type I):
In both mechanisms, particles gain energy from compressible wave motions via reflection by the longitudinal magnetic field gradients 
\citep[e.g.,][]{Fermi_1949,Brunetti:2007zp, Demidem:2019jzn,Huang_2024}.
We note that the term ``mirror acceleration" used in some literature in fact refers to the TTD
\citep[e.g.,][]{Vega:2024pkx}.
The gradient-B drift acceleration has important applications to shock and reconnection acceleration 
\citep{Sonnerup_1969,Xu_2022,Xu_2023}.
It usually requires the drift motion to be slower than the gyromotion, i.e., a particle's orbit intersects the magnetic field gradient many times, 
and $J_1$ is conserved 
\citep{Drury_1983,Kirk_1989}.
Both TTD and gradient-B drift acceleration
increase $p_\|$. 
Consequently, 
the longitudinal mirroring condition would soon be violated by the acceleration itself, and thus the acceleration cannot be self-sustained. 
The accelerated particles would also escape 
along the magnetic field and cannot be spatially confined to undergo repeated acceleration
toward higher energies.

(3) Curvature drift acceleration (Type I): 
Particles following a curved moving field line can gain energy from its turbulent motion
\citep{Fermi_1954}.
The curvature drift acceleration increases $p_\|$. 
It usually requires the drift motion to be slower than the gyromotion, 
and $J_1$ is conserved.
High curvature is usually found to correlate and  
co-locate with weak magnetic fields \citep[e.g.,][]{Yang_2019}.
Therefore, the curvature drift acceleration is expected to primarily occur in the weak-field regions and the reconnection regions in turbulence.
We also note that the guiding-center framework for analyzing the curvature drift acceleration  \citep[see, e.g.,][]{Sebastian:2025dex}
does not apply to the acceleration of energetic particles that we focus on here. 
They experience a fast variation of the magnetic field and $J_1$ within one gyro-orbit.

{(4) Betatron acceleration (Type II):}
As a Type II mechanism, betatron acceleration is due to the temporal variations of magnetic field strengths. 
In nonrelativistic turbulence, 
it is formulated under the consideration of fast particle gyrations in a slowly varying magnetic field with $J_1$ conserved.  
An increase in $p_\perp$ is thus linked with a decrease in $r_L$. 
The process is reversible.  
If the varying magnetic field has a constant averaged value, the particle energy also remains constant on average
\citep{Ginzburg_1964}. 

(5) Magnetic pumping (Type II):
Magnetic pumping is traditionally considered as a particle heating mechanism \citep[e.g.,][]{Lichko_2017}.
In nonrelativistic turbulence, similar to the betatron acceleration, 
it is based on the conservation of $J_1$ and reversible change of a particle's $p_\perp$. 
For net acceleration to occur, the key is pitch-angle scattering, which breaks reversibility and transfers momentum to the parallel direction 
\citep{Malkov_2026}.
However, by invoking efficient scattering, it becomes subdominant to Type I scattering acceleration. 


{(6) Mirror acceleration (Type II):}
As a mechanism established based on the improved understanding of turbulent magnetic fields 
\citep{Goldreich:1995sr,Lazarian:1998wd},
the mirror acceleration 
enables both stochasticity and efficiency of Type II acceleration, 
as the two major difficulties faced by betatron acceleration and magnetic pumping. 
Regardless of the plasma $\beta$ and the plasma compressibility, as long as there are compressions of magnetic fields that cause the spatial and temporal variations of magnetic field strength, the mirror acceleration is expected. 
In the compressible plasma as simulated here, 
fast and slow modes 
\citep{Takamoto:2016kdu,Takamoto:2017vhf, Sebastian:2026fwz} 
compress magnetic fields and contribute to the mirror acceleration.

In nonrelativistic turbulence,
particles sample the magnetic flux change via 
their perpendicular superdiffusion.
They stochastically encounter different compressed/expanded turbulent eddies, 
which breaks reversibility and enables stochasticity in acceleration. 
The mirrors regulating the parallel mirror diffusion can be much smaller than the eddies dominating the particle acceleration. 
The latter have the 
timescale of turbulent compression 
comparable to the mirror diffusion time of  particles (LX23).

In relativistic turbulence, significant acceleration can happen during one mirror interaction. 
A particle samples both spatial and temporal variations of the magnetic field 
via its gyration. 
Its perpendicular momentum increases due to the rapid change in magnetic flux via the relativistic turbulent compression of the magnetic field.  
Both the intrinsic perpendicular superdiffuion 
of Alfvénic turbulence and the distorted gyro-orbits allow particles to spatially sample different compressed/expanded turbulent eddies and thus enable stochasticity.


In brief,  
Type I mechanisms
preferentially accelerate particles parallel to the magnetic field, while 
Type II mechanisms  
preferentially accelerate particles perpendicular to the magnetic field.  
The mirror acceleration resolves the major theoretical difficulties of traditional Type II mechanisms and satisfactorily explains the 
anisotropic distribution of accelerated particles concentrated at large pitch angles, 
which cannot be explained by Type I mechanisms. 
Our result suggests 
the dominance of the mirror acceleration 
compared to other candidate mechanisms in accelerating particles to high energies.



\subsection{2D vs. 3D and driven vs. decaying turbulence}

The theory of mirror acceleration (LX23) is established for 3D turbulence. 
We note that anisotropic distributions of accelerated particles have been reported in both 2D and 3D PIC simulations with decaying turbulence \citep[e.g.,][]{Comisso:2018kuh, Comisso:2019frj}.
In 2D, without perpendicular superdiffusion of turbulent magnetic fields,
particles are not allowed to spatially access 
different compressed/expanded turbulent eddies. 
They are trapped within 2D magnetic islands or between two merging islands. 
If an island undergoes oscillatory contraction and expansion, 
the acceleration is reversible, with zero net energy gain. 
If an island only undergoes contraction, 
the acceleration
stops when the contraction ends. 
Therefore, both the acceleration efficiency and the maximum 
energy are limited compared to the acceleration in 3D. 
Their comparison should be performed with a sufficiently large numerical domain, especially for the 3D simulation. 
In addition, 
compared to the long-lived 2D magnetic islands, 3D magnetized turbulence decays rapidly. Therefore, 
continuous driving should be applied for the comparison of turbulent acceleration in 2D and 3D. 

Compared with driven turbulence, 
decaying turbulence evolves toward a magnetic force-free state, with reconnection-dominated magnetic structures \citep[e.g.,][]{Hosking:2020wom, Dong:2022crn}.
With fundamentally different dynamics and magnetic structures, 
the particle acceleration in decaying turbulence, especially in the relaxed state, 
is expected to differ significantly from that in driven turbulence.

\subsection{Energy spectrum of accelerated particles}
It is known that the competition between acceleration and escape of particles naturally results in a power-law energy distribution of the accelerated particles
\citep{Longair_2011}. 
A non-universal power-law slope has been reported in earlier studies, 
with a harder slope found for  
a higher magnetization, a stronger turbulence level, and driven turbulence in comparison with decaying turbulence 
\citep{Comisso:2018kuh,Zhdankin:2016lta}.
Naturally, these favorable conditions for efficient acceleration act to harden the resulting particle energy spectrum. 
A direct comparison between the 
simulated spectrum that evolves in time, and the 
analytical solution for a steady-state spectrum 
(e.g., \cite{Xu:2017ypi})
is tricky. 
Nevertheless,
as the mirror acceleration suppresses the escape of particles 
from the acceleration site
by increasing their pitch angles, 
a hard spectrum with the slope approaching $-1$ is generally 
expected as a result of the mirror acceleration.

\subsection{Astrophysical implications}

With a limited numerical resolution, the inertial range of turbulence and thus the energy range of accelerated particles studied in this work are limited. Further numerical tests and studies with higher-resolution PIC simulations, and their comparisons with relativistic MHD simulations, are necessary and useful for astrophysical applications.

The mirror acceleration results in 
a more anisotropic pitch angle distribution of more energetic particles.
This is consistent with the recent observational finding on a higher polarization degree at a higher frequency, 
based on radio to X-ray polarization measurements of blazars \citep[e.g.,][]{2022Natur.611..677L, MAGIC:2024tyq}.
Moreover, the anisotropic distribution of accelerated electrons can affect their synchrotron spectral shape 
\citep[e.g.,][]{Yang:2018zrd, Comisso:2021isw, Comisso:2024iyx}. 
These observational features 
can be used to diagnose and test the mirror acceleration in high-energy astrophysical sources.

The mirror acceleration also applies to protons. Its numerical testing with PIC simulations in an electron-proton plasma 
will be carried out in our future work. 
Further studies on the efficiency of mirror acceleration in energizing protons will have important implications for explaining observations of (ultra) high-energy cosmic neutrinos
\citep{Das:2025vqd}.

\section{Conclusions \label{sec:summary}}

We perform the first numerical study on Type II mirror acceleration mechanism in relativistic turbulence with a 3D PIC simulation of pair plasma. 
Turbulent compressions of magnetic fields naturally create 
temporal variations of magnetic field strengths,
entailing the mirror acceleration and a non-thermal population of particles. 
Our main findings are as follows. 

\begin{figure*}
\centering
\subfigure[]{
   \includegraphics[width=0.33\textwidth]{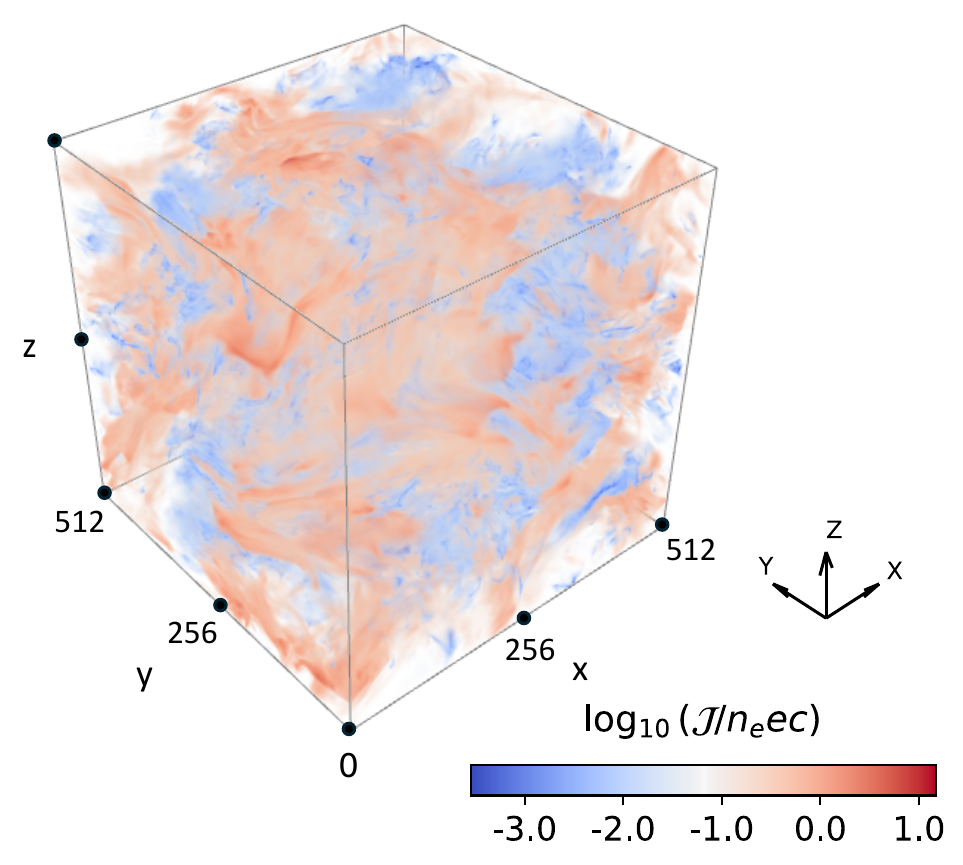}\label{fig:j11}}%
\subfigure[]{
   \includegraphics[width=0.33\textwidth]{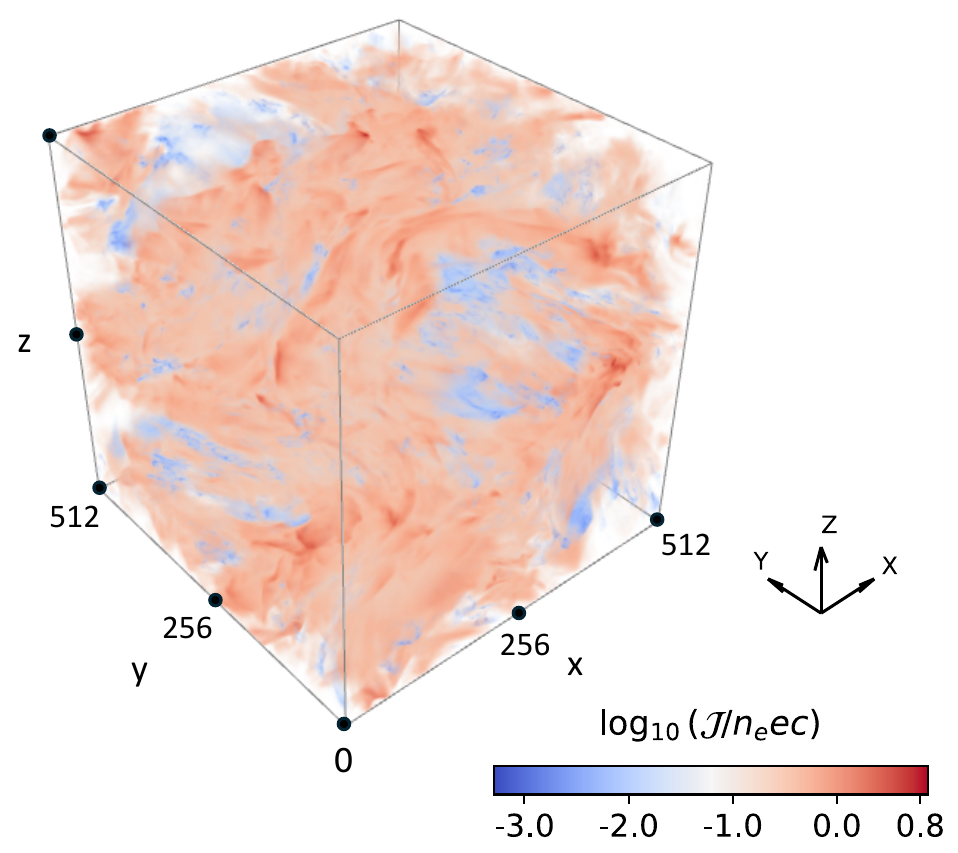}\label{fig:j14}}%
\subfigure[]{
\includegraphics[width=0.33\textwidth]{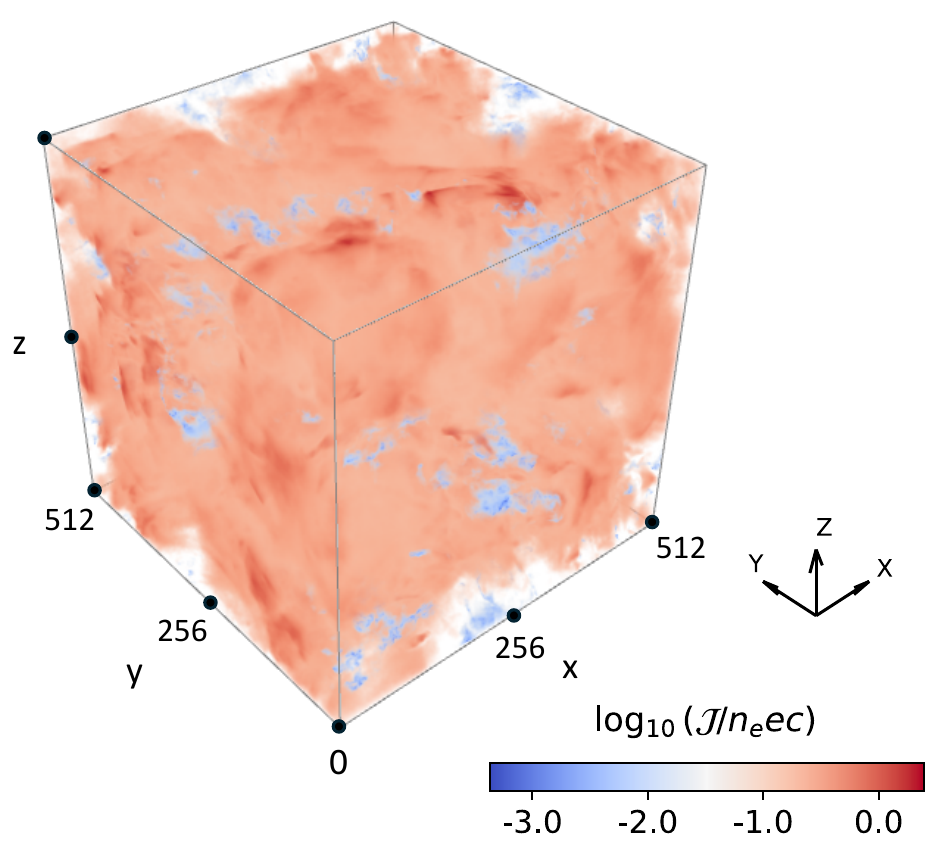}\label{fig:j22}}
\caption{Distribution of $\mathcal{J}$ (normalized by $n_e e c$) measured at (a) $t=1.1\ell_0/c$, (b) $t=1.4\ell_0/c$, and (c) $t=2.2\ell_0/c$.}
\label{fig:jdist}
\end{figure*}

\begin{figure}
\vspace*{0.5cm}
    \centering
    \includegraphics[width=0.46\textwidth]{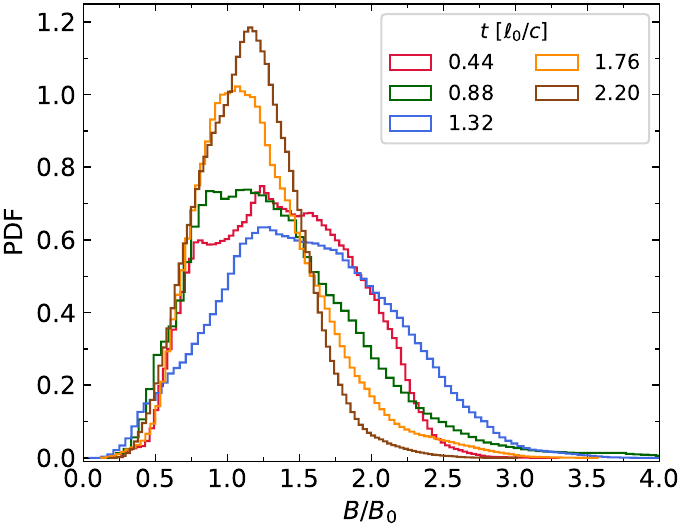}
    \caption{PDFs of $B$ measured at different times.}
    \label{fig:B_dist}
\end{figure}


\begin{enumerate}[leftmargin=*]
  \item 
The 
acceleration of high-energy particles in the simulation
is dominated by the mirror acceleration, characterized by a stochastic increase in perpendicular momentum and pitch angle of the particles. 
The fractional increase in perpendicular momentum is positively correlated with the local strengthening of compressed magnetic fields. 
This feature distinguishes this mechanism from Type I mechanisms, e.g., scattering acceleration, TTD, curvature drift acceleration, that primarily increase the parallel momentum.  

\item
The energized particles via the mirror acceleration are well confined in space due to their large pitch angles, 
facilitating efficient and self-sustained further mirror acceleration. 
The characteristic anisotropic particle distribution concentrated at large pitch angles 
has important observational implications on multi-frequency polarization measurements and synchrotron spectral features. 



\item
Compared with the mirror acceleration in nonrelativistic turbulence (LX23),
interactions with transverse mirrors are more favored for the mirror acceleration in relativistic turbulence. 
Reflections by longitudinal mirrors, which are traditionally considered for mirror interactions, become irrelevant for the acceleration. 
In relativistic turbulence, 
significant magnetic flux change and thus particle energy change can happen during one mirror interaction within one gyro-orbit.
This causes a systematic change of the first adiabatic invariant during the mirror interaction 
and distorted gyro-orbits. 

\item
The reconnection acceleration of particles interacting with local weakening fields
occurs when $r_L$ is smaller than the current layer thickness at an early time of the simulation. 
It can result in 
{a boost in perpendicular momentum
and serves as a pre-acceleration process 
for the subsequent mirror acceleration.}
In the absence of cooling, the accelerated particles more frequently encounter the  
growing
reconnection regions toward the end of the simulation. 
However, this is not expected in reality with well-separated gyroradii of non-thermal and thermal particles.

\item
One must exercise great caution when comparing turbulent acceleration in 2D and 3D, and in driven and decaying turbulence. 
In the absence of turbulence stochasticity in 2D, acceleration is expected to be inefficient, 
but the relaxation of 2D magnetic structures is much slower than that of 3D structures. 
The combination of these effects may make turbulent acceleration appear similar in 2D and 3D
\citep[e.g.,][]{Comisso:2018kuh, Comisso:2019frj}. 
However, the underlying physics is fundamentally different. 
Their comparison should be performed with driven turbulence.

    %
    %
\end{enumerate}

These findings suggest that the mirror acceleration can play a significant role in energizing {high-energy}
particles 
in {turbulent} astrophysical environments, motivating its further studies using high-resolution electron-proton plasma simulations.

\begin{acknowledgments}
S.D. and S.X. acknowledge the support from the NASA ATP award 80NSSC24K0896
and the NASA Heliophysics Living with A Star Science Program
80NSSC25K0067. The authors acknowledge UFIT Research Computing for providing computational resources and support contributing to the research results reported in this work. 
\end{acknowledgments}


\appendix


\section{Transformation to comoving frame \label{app:drift}}
We define the drift velocity as \citep{Takeuchi_2002},
\begin{equation}
\begin{split}
    \bm{\beta_d} = \begin{cases}
        \bm{E\times B}/B^2 \ \ \ \ (E<B),\\
        \bm{E\times B}/E^2 \ \ \ \ (E>B).
    \end{cases}
    \label{eqn:frame}
\end{split}
\end{equation}
where $\beta$ is the velocity normalized by c. The two equations are introduced so that $\bm{\beta_d}<1$. By boosting the particle velocity from the laboratory frame to the frame moving with $\bm{v_d}$,
we have 
\begin{equation}
\begin{split}
    \boldsymbol{\beta}^\prime = \frac{1}{1-\boldsymbol{\beta\cdot\beta_d}}\left[\frac{\boldsymbol{\beta}}{\Gamma_d} -\boldsymbol{\beta_d}+\frac{\Gamma_d}{1+\Gamma_d}(\boldsymbol{\beta\cdot\beta_d})\boldsymbol{\beta_d}\right],
\end{split}
\end{equation}
where $\bm{\beta_d}=\boldsymbol{v_d}/c$, $\Gamma_d=1/\sqrt{1-\beta_d^2}$,  $\boldsymbol{\beta}=\boldsymbol{v}/c$, 
$\boldsymbol{\beta^\prime}=\boldsymbol{v^\prime}/c$, 
and $\boldsymbol{v}$ and $\boldsymbol{v^\prime}$ are the particle velocity in the laboratory and comoving frame, respectively. The particle Lorentz factor transforms as 
\begin{equation}
\begin{split}
\gamma_e^\prime = \gamma_e \Gamma_d \left( 1 - \boldsymbol{\beta} \cdot \boldsymbol{\beta}_d \right),
\end{split}
\end{equation}
where $\gamma_e= 1/\sqrt{1-\beta^2}$ and $\gamma_e^\prime = 1/\sqrt{1-\beta^{\prime 2}}$. By measuring the quantities in the frames moving with $\bm{v_d}$, the energy oscillations associated with particle gyrations seen in the laboratory frame are removed \citep[e.g.,][]{Bresci_2022, Comisso:2019frj}.

\section{Temporal evolution of 
distributions of  current density 
and magnetic field strength\label{app:jdist}}

Fig.~\ref{fig:jdist} illustrates the distribution of current density $\mathcal{J}$. 
 In the absence of cooling, 
 with the increase of gyroradius of thermal particles,
 we clearly see the thickening of current layers with time.
 As the reconnection regions 
 with weakening magnetic fields
 become more and more volume-filling, 
 in Fig.~\ref{fig:B_dist}, we see that the PDF of $B$ narrows with time, and its peak moves toward lower $B$ values.
\software{RUNKO \citep{2022A&A...664A..68N}
          }



\bibliography{sample631}{}

@ARTICLE{2022A&A...664A..68N,
       author = {{N{\"a}ttil{\"a}}, J.},
        title = "{Runko: Modern multiphysics toolbox for plasma simulations}",
      journal = {Astron. Astrophys.},
         year = 2022,
        month = aug,
       volume = {664},
          eid = {A68},
        pages = {A68},
          doi = {10.1051/0004-6361/201937402},
archivePrefix = {arXiv},
       eprint = {1906.06306},
 primaryClass = {physics.comp-ph},
       adsurl = {https://ui.adsabs.harvard.edu/abs/2022A&A...664A..68N}
}

@ARTICLE{nattila2024,
       author = {{N{\"a}ttil{\"a}}, Joonas},
        title = "{Radiative plasma simulations of black hole accretion flow coronae in the hard and soft states}",
      journal = {Nature Communications},
         year = 2024,
        month = dec,
       volume = {15},
       number = {1},
          eid = {7026},
        pages = {7026},
          doi = {10.1038/s41467-024-51257-1},
archivePrefix = {arXiv},
       eprint = {2408.08161},
 primaryClass = {astro-ph.HE},
       adsurl = {https://ui.adsabs.harvard.edu/abs/2024NatCo..15.7026N}
}

@article{Nattila:2021qag,
    author = {N\"attil\"a, Joonas and Beloborodov, Andrei M.},
    title = "{Heating of Magnetically Dominated Plasma by Alfv\'en-Wave Turbulence}",
    eprint = "2111.15578",
    archivePrefix = "arXiv",
    primaryClass = "astro-ph.HE",
    doi = "10.1103/PhysRevLett.128.075101",
    journal = "Phys. Rev. Lett.",
    volume = "128",
    number = "7",
    pages = "075101",
    year = "2022"
}

@ARTICLE{vega2022,
       author = {{Vega}, Cristian and {Boldyrev}, Stanislav and {Roytershteyn}, Vadim and {Medvedev}, Mikhail},
        title = "{Turbulence and Particle Acceleration in a Relativistic Plasma}",
      journal = {Astrophys. J. Lett.},
         year = 2022,
        month = jan,
       volume = {924},
       number = {1},
          eid = {L19},
        pages = {L19},
          doi = {10.3847/2041-8213/ac441e},
archivePrefix = {arXiv},
       eprint = {2111.04907},
 primaryClass = {physics.plasm-ph},
       adsurl = {https://ui.adsabs.harvard.edu/abs/2022ApJ...924L..19V}
}

@ARTICLE{vega2024,
       author = {{Vega}, Cristian and {Boldyrev}, Stanislav and {Roytershteyn}, Vadim},
        title = "{Particle Acceleration in Relativistic Alfv{\'e}nic Turbulence}",
      journal = {Astrophys. J.},
         year = 2024,
        month = aug,
       volume = {971},
       number = {1},
          eid = {106},
        pages = {106},
          doi = {10.3847/1538-4357/ad5f8f},
archivePrefix = {arXiv},
       eprint = {2405.07891},
 primaryClass = {physics.plasm-ph},
       adsurl = {https://ui.adsabs.harvard.edu/abs/2024ApJ...971..106V}
}

@ARTICLE{meringolo2023,
       author = {{Meringolo}, Claudio and {Cruz-Osorio}, Alejandro and {Rezzolla}, Luciano and {Servidio}, Sergio},
        title = "{Microphysical Plasma Relations from Special-relativistic Turbulence}",
      journal = {Astrophys. J.},
         year = 2023,
        month = feb,
       volume = {944},
       number = {2},
          eid = {122},
        pages = {122},
          doi = {10.3847/1538-4357/acaefe},
archivePrefix = {arXiv},
       eprint = {2301.02669},
 primaryClass = {astro-ph.HE},
       adsurl = {https://ui.adsabs.harvard.edu/abs/2023ApJ...944..122M}
}

@ARTICLE{lemoine2025,
       author = {{Lemoine}, M.},
        title = "{Effective theory for stochastic particle acceleration, with application to magnetized turbulence}",
      journal = {arXiv e-prints},
         year = 2025,
        month = jan,
          eid = {arXiv:2501.19136},
        pages = {arXiv:2501.19136},
          doi = {10.48550/arXiv.2501.19136},
archivePrefix = {arXiv},
       eprint = {2501.19136},
 primaryClass = {physics.plasm-ph},
       adsurl = {https://ui.adsabs.harvard.edu/abs/2025arXiv250119136L}
}

@ARTICLE{ha2024,
       author = {{Ha}, Trung and {N{\"a}ttil{\"a}}, Joonas and {Davelaar}, Jordy and {Sironi}, Lorenzo},
        title = "{Machine-Learning Characterization of Intermittency in Plasma Turbulence: Single and Double Sheet Structures}",
      journal = {arXiv e-prints},
         year = 2024,
        month = oct,
          eid = {arXiv:2410.01878},
        pages = {arXiv:2410.01878},
          doi = {10.48550/arXiv.2410.01878},
archivePrefix = {arXiv},
       eprint = {2410.01878},
 primaryClass = {astro-ph.HE},
       adsurl = {https://ui.adsabs.harvard.edu/abs/2024arXiv241001878H}
}

@article{Lazarian:2021kvd,
    author = "Lazarian, Alex and Xu, Siyao",
    title = "{Diffusion of Cosmic Rays in MHD Turbulence with Magnetic Mirrors}",
    eprint = "2106.08362",
    archivePrefix = "arXiv",
    primaryClass = "astro-ph.HE",
    doi = "10.3847/1538-4357/ac2de9",
    journal = "Astrophys. J.",
    volume = "923",
    number = "1",
    pages = "53",
    year = "2021"
}

@article{Kirk:2001zv,
    author = "Kirk, J. G. and Dendy, R. O.",
    editor = "Giller, M.",
    title = "{Shock acceleration of cosmic rays: A Critical review}",
    eprint = "astro-ph/0101175",
    archivePrefix = "arXiv",
    doi = "10.1088/0954-3899/27/7/316",
    journal = "J. Phys. G",
    volume = "27",
    pages = "1589--1596",
    year = "2001"
}

@article{Blasi:2013rva,
    author = "Blasi, Pasquale",
    title = "{The Origin of Galactic Cosmic Rays}",
    eprint = "1311.7346",
    archivePrefix = "arXiv",
    primaryClass = "astro-ph.HE",
    doi = "10.1007/s00159-013-0070-7",
    journal = "Astron. Astrophys. Rev.",
    volume = "21",
    pages = "70",
    year = "2013"
}

@ARTICLE{2012SSRv..173..309B,
       author = {{Bykov}, Andrei and {Gehrels}, Neil and {Krawczynski}, Henric and {Lemoine}, Martin and {Pelletier}, Guy and {Pohl}, Martin},
        title = "{Particle Acceleration in Relativistic Outflows}",
      journal = {Space Sci. Rev.},
         year = 2012,
        month = nov,
       volume = {173},
       number = {1-4},
        pages = {309-339},
          doi = {10.1007/s11214-012-9896-y},
archivePrefix = {arXiv},
       eprint = {1205.2208},
 primaryClass = {astro-ph.HE},
       adsurl = {https://ui.adsabs.harvard.edu/abs/2012SSRv..173..309B}
}

@article{Kowal:2012rv,
    author = "Kowal, Grzegorz and de Gouveia Dal Pino, Elisabete M. and Lazarian, A.",
    title = "{Particle Acceleration in Turbulence and Weakly Stochastic Reconnection}",
    eprint = "1202.5256",
    archivePrefix = "arXiv",
    primaryClass = "astro-ph.HE",
    doi = "10.1103/PhysRevLett.108.241102",
    journal = "Phys. Rev. Lett.",
    volume = "108",
    pages = "241102",
    year = "2012"
}

@article{Meszaros:2019xej,
    author = "M\'esz\'aros, P\'eter and Fox, Derek B. and Hanna, Chad and Murase, Kohta",
    title = "{Multi-Messenger Astrophysics}",
    eprint = "1906.10212",
    archivePrefix = "arXiv",
    primaryClass = "astro-ph.HE",
    doi = "10.1038/s42254-019-0101-z",
    journal = "Nature Rev. Phys.",
    volume = "1",
    pages = "585--599",
    year = "2019"
}

@article{Zhdankin:2016lta,
    author = "Zhdankin, Vladimir and Werner, Gregory R. and Uzdensky, Dmitri A. and Begelman, Mitchell C.",
    title = "{Kinetic Turbulence in Relativistic Plasma: From Thermal Bath to Nonthermal Continuum}",
    eprint = "1609.04851",
    archivePrefix = "arXiv",
    primaryClass = "physics.plasm-ph",
    doi = "10.1103/PhysRevLett.118.055103",
    journal = "Phys. Rev. Lett.",
    volume = "118",
    number = "5",
    pages = "055103",
    year = "2017"
}

@article{Zhdankin:2019dfz,
    author = "Zhdankin, Vladimir and Uzdensky, Dmitri A. and Werner, Gregory R. and Begelman, Mitchell C.",
    title = "{Kinetic turbulence in shining pair plasma: intermittent beaming and thermalization by radiative cooling}",
    eprint = "1908.08032",
    archivePrefix = "arXiv",
    primaryClass = "astro-ph.HE",
    doi = "10.1093/mnras/staa284",
    journal = "Mon. Not. Roy. Astron. Soc.",
    volume = "493",
    number = "1",
    pages = "603--626",
    year = "2020"
}

@article{Comisso:2018kuh,
    author = "Comisso, Luca and Sironi, Lorenzo",
    title = "{Particle Acceleration in Relativistic Plasma Turbulence}",
    eprint = "1809.01168",
    archivePrefix = "arXiv",
    primaryClass = "astro-ph.HE",
    doi = "10.1103/PhysRevLett.121.255101",
    journal = "Phys. Rev. Lett.",
    volume = "121",
    number = "25",
    pages = "255101",
    year = "2018"
}

@article{Comisso:2019frj,
    author = "Comisso, Luca and Sironi, Lorenzo",
    title = "{The interplay of magnetically-dominated turbulence and magnetic reconnection in producing nonthermal particles}",
    eprint = "1909.01420",
    archivePrefix = "arXiv",
    primaryClass = "astro-ph.HE",
    doi = "10.3847/1538-4357/ab4c33",
    journal = "Astrophys. J.",
    volume = "886",
    pages = "122",
    year = "2019"
}

@article{Fermi_1949,
  title = {On the Origin of the Cosmic Radiation},
  author = {Fermi, ENRICO},
  journal = {Phys. Rev.},
  volume = {75},
  issue = {8},
  pages = {1169--1174},
  numpages = {0},
  year = {1949},
  month = {Apr},
  publisher = {American Physical Society},
  doi = {10.1103/PhysRev.75.1169},
  url = {https://link.aps.org/doi/10.1103/PhysRev.75.1169}
}

@article{Wong:2019dog,
    author = "Wong, Kai and Zhdankin, Vladimir and Uzdensky, Dmitri A. and Werner, Gregory R. and Begelman, Mitchell C.",
    title = "{First-principles demonstration of diffusive-advective particle acceleration in kinetic simulations of relativistic plasma turbulence}",
    eprint = "1901.03439",
    archivePrefix = "arXiv",
    primaryClass = "astro-ph.HE",
    doi = "10.3847/2041-8213/ab8122",
    journal = "Astrophys. J. Lett.",
    volume = "893",
    number = "1",
    pages = "L7",
    year = "2020"
}

@article{Comisso:2024iyx,
    author = "Comisso, Luca",
    title = "{Concurrent Particle Acceleration and Pitch-angle Anisotropy Driven by Magnetic Reconnection: Ion-electron Plasmas}",
    eprint = "2405.18227",
    archivePrefix = "arXiv",
    primaryClass = "astro-ph.HE",
    doi = "10.3847/1538-4357/ad51fe",
    journal = "Astrophys. J.",
    volume = "972",
    number = "1",
    pages = "9",
    year = "2024"
}

@article{Zhdankin:2018doa,
    author = "Zhdankin, Vladimir and Uzdensky, Dmitri A. and Werner, Gregory R. and Begelman, Mitchell C.",
    title = "{System-size convergence of nonthermal particle acceleration in relativistic plasma turbulence}",
    eprint = "1805.08754",
    archivePrefix = "arXiv",
    primaryClass = "astro-ph.HE",
    doi = "10.3847/2041-8213/aae88c",
    journal = "Astrophys. J. Lett.",
    volume = "867",
    number = "1",
    pages = "L18",
    year = "2018"
}

@article{Lazarian:2023zui,
    author = "Lazarian, Alex and Xu, Siyao",
    title = "{Mirror Acceleration of Cosmic Rays in a High-\ensuremath{\beta} Medium}",
    eprint = "2306.14973",
    archivePrefix = "arXiv",
    primaryClass = "astro-ph.HE",
    doi = "10.3847/1538-4357/acea5c",
    journal = "Astrophys. J.",
    volume = "956",
    number = "1",
    pages = "63",
    year = "2023"
}

@article{Comisso:2021isw,
    author = "Comisso, Luca and Sironi, Lorenzo",
    title = "{Pitch-Angle Anisotropy Controls Particle Acceleration and Cooling in Radiative Relativistic Plasma Turbulence}",
    eprint = "2109.02666",
    archivePrefix = "arXiv",
    primaryClass = "physics.plasm-ph",
    doi = "10.1103/PhysRevLett.127.255102",
    journal = "Phys. Rev. Lett.",
    volume = "127",
    number = "25",
    pages = "255102",
    year = "2021"
}

@ARTICLE{Jokipii_1966,
       author = {{Jokipii}, J.~R.},
        title = "{Cosmic-Ray Propagation. I. Charged Particles in a Random Magnetic Field}",
      journal = {Astrophys. J.},
         year = 1966,
        month = nov,
       volume = {146},
        pages = {480},
          doi = {10.1086/148912},
       adsurl = {https://ui.adsabs.harvard.edu/abs/1966ApJ...146..480J}
}

@article{Zhang:2023igz,
    author = "Zhang, Chao and Xu, Siyao",
    title = "{Numerical Testing of Mirror Diffusion of Cosmic Rays}",
    eprint = "2311.18001",
    archivePrefix = "arXiv",
    primaryClass = "astro-ph.HE",
    doi = "10.3847/2041-8213/ad0fe5",
    journal = "Astrophys. J. Lett.",
    volume = "959",
    number = "1",
    pages = "L8",
    year = "2023"
}

@article{Murase:2019tjj,
    author = "Murase, Kohta and Bartos, Imre",
    title = "{High-Energy Multimessenger Transient Astrophysics}",
    eprint = "1907.12506",
    archivePrefix = "arXiv",
    primaryClass = "astro-ph.HE",
    doi = "10.1146/annurev-nucl-101918-023510",
    journal = "Ann. Rev. Nucl. Part. Sci.",
    volume = "69",
    pages = "477--506",
    year = "2019"
}

@article{Lemoine:2021mtv,
    author = "Lemoine, Martin",
    title = "{Particle acceleration in strong MHD turbulence}",
    eprint = "2104.08199",
    archivePrefix = "arXiv",
    primaryClass = "astro-ph.HE",
    doi = "10.1103/PhysRevD.104.063020",
    journal = "Phys. Rev. D",
    volume = "104",
    number = "6",
    pages = "063020",
    year = "2021"
}

@article{Bresci_2022,
  title = {Nonresonant particle acceleration in strong turbulence: Comparison to kinetic and MHD simulations},
  author = {Bresci, Virginia and Lemoine, Martin and Gremillet, Laurent and Comisso, Luca and Sironi, Lorenzo and Demidem, Camilia},
  journal = {Phys. Rev. D},
  volume = {106},
  issue = {2},
  pages = {023028},
  numpages = {17},
  year = {2022},
  month = {Jul},
  publisher = {American Physical Society},
  doi = {10.1103/PhysRevD.106.023028},
  url = {https://link.aps.org/doi/10.1103/PhysRevD.106.023028}
}

@ARTICLE{Fermi_1954,
       author = {{Fermi}, E.},
        title = "{Galactic Magnetic Fields and the Origin of Cosmic Radiation.}",
      journal = {Astrophys. J.},
         year = 1954,
        month = jan,
       volume = {119},
        pages = {1},
          doi = {10.1086/145789},
       adsurl = {https://ui.adsabs.harvard.edu/abs/1954ApJ...119....1F}
}

@article{Comisso:2022iqy,
    author = "Comisso, Luca and Sironi, Lorenzo",
    title = "{Ion and Electron Acceleration in Fully Kinetic Plasma Turbulence}",
    eprint = "2209.04475",
    archivePrefix = "arXiv",
    primaryClass = "astro-ph.HE",
    doi = "10.3847/2041-8213/ac8422",
    journal = "Astrophys. J. Lett.",
    volume = "936",
    number = "2",
    pages = "L27",
    year = "2022"
}

@article{Kempski:2023ikw,
    author = "Kempski, Philipp and Fielding, Drummond B. and Quataert, Eliot and Galishnikova, Alisa K. and Kunz, Matthew W. and Philippov, Alexander A. and Ripperda, Bart",
    title = "{Cosmic ray transport in large-amplitude turbulence with small-scale field reversals}",
    eprint = "2304.12335",
    archivePrefix = "arXiv",
    primaryClass = "astro-ph.HE",
    doi = "10.1093/mnras/stad2609",
    journal = "Mon. Not. Roy. Astron. Soc.",
    volume = "525",
    number = "4",
    pages = "4985--4998",
    year = "2023"
}

@article{Groselj:2023bgy,
    author = "Groselj, Daniel and Hakobyan, Hayk and Beloborodov, Andrei M. and Sironi, Lorenzo and Philippov, Alexander",
    title = "{Radiative Particle-in-Cell Simulations of Turbulent Comptonization in Magnetized Black-Hole Coronae}",
    eprint = "2301.11327",
    archivePrefix = "arXiv",
    primaryClass = "astro-ph.HE",
    doi = "10.1103/PhysRevLett.132.085202",
    journal = "Phys. Rev. Lett.",
    volume = "132",
    number = "8",
    pages = "085202",
    year = "2024"
}

@ARTICLE{Kulsrud_1969,
       author = {{Kulsrud}, Russell and {Pearce}, William P.},
        title = "{The Effect of Wave-Particle Interactions on the Propagation of Cosmic Rays}",
      journal = {Astrophys. J.},
         year = 1969,
        month = may,
       volume = {156},
        pages = {445},
          doi = {10.1086/149981},
       adsurl = {https://ui.adsabs.harvard.edu/abs/1969ApJ...156..445K}
}

@article{Petrosian:2004ft,
    author = "Petrosian, Vahe and Liu, Si-Ming",
    title = "{Stochastic acceleration of electrons and protons. 1. Acceleration by parallel propagating waves}",
    eprint = "astro-ph/0401585",
    archivePrefix = "arXiv",
    doi = "10.1086/421486",
    journal = "Astrophys. J.",
    volume = "610",
    pages = "550--571",
    year = "2004"
}

@article{Brunetti:2007zp,
    author = "Brunetti, Gianfranco and Lazarian, A.",
    title = "{Compressible Turbulence in Galaxy Clusters: Physics and Stochastic Particle Re-acceleration}",
    eprint = "astro-ph/0703591",
    archivePrefix = "arXiv",
    doi = "10.1111/j.1365-2966.2007.11771.x",
    journal = "Mon. Not. Roy. Astron. Soc.",
    volume = "378",
    pages = "245--275",
    year = "2007"
}

@article{Zhang:2024evq,
    author = "Zhang, Chao and Xu, Siyao",
    title = "{Cosmic Ray Diffusion in Magnetic Fields Amplified by Nonlinear Turbulent Dynamo}",
    eprint = "2406.03542",
    archivePrefix = "arXiv",
    primaryClass = "astro-ph.HE",
    doi = "10.3847/1538-4357/ad79fb",
    journal = "Astrophys. J.",
    volume = "975",
    number = "1",
    pages = "65",
    year = "2024"
}

@article{Barreto-Mota:2024kli,
    author = "Barreto-Mota, Lucas and de Gouveia Dal Pino, Elisabete M. and Xu, Siyao and Lazarian, Alexandre",
    title = "{Cosmic-Ray Diffusion in the Turbulent Interstellar Medium: Effects of Mirror Diffusion and Pitch-angle Scattering}",
    eprint = "2405.12146",
    archivePrefix = "arXiv",
    primaryClass = "astro-ph.HE",
    doi = "10.3847/1538-4357/ade4c8",
    journal = "Astrophys. J.",
    volume = "988",
    number = "2",
    pages = "269",
    year = "2025"
}

@article{Fiorillo:2024akm,
    author = "Fiorillo, Damiano F. G. and Comisso, Luca and Peretti, Enrico and Petropoulou, Maria and Sironi, Lorenzo",
    title = "{A Magnetized Strongly Turbulent Corona as the Source of Neutrinos from NGC 1068}",
    eprint = "2407.01678",
    archivePrefix = "arXiv",
    primaryClass = "astro-ph.HE",
    doi = "10.3847/1538-4357/ad7021",
    journal = "Astrophys. J.",
    volume = "974",
    number = "1",
    pages = "75",
    year = "2024"
}

@ARTICLE{Cesarksky_1973,
       author = {{Cesarsky}, Catherine J. and {Kulsrud}, Russell M.},
        title = "{Role of Hydromagnetic Waves in Cosmic-Ray Confinement in the Disk. Theory of Behavior in General Wave Spectra}",
      journal = {Astrophys. J.},
         year = 1973,
        month = oct,
       volume = {185},
        pages = {153-166},
          doi = {10.1086/152405},
       adsurl = {https://ui.adsabs.harvard.edu/abs/1973ApJ...185..153C}
}

@article{Xiao:2025yrt,
    author = "Xiao, Ya-Wen and Zhang, Jian-Fu and Xu, Siyao",
    title = "{Studying the diffusion mechanism of cosmic-ray particles}",
    eprint = "2506.15031",
    archivePrefix = "arXiv",
    primaryClass = "astro-ph.HE",
    doi = "10.1051/0004-6361/202453340",
    journal = "Astron. Astrophys.",
    volume = "699",
    pages = "A317",
    year = "2025"
}

@article{Nattila:2020qfx,
    author = {N\"attil\"a, Joonas and Beloborodov, Andrei M.},
    title = "{Radiative Turbulent Flares in Magnetically Dominated Plasmas}",
    eprint = "2012.03043",
    archivePrefix = "arXiv",
    primaryClass = "astro-ph.HE",
    doi = "10.3847/1538-4357/ac1c76",
    journal = "Astrophys. J.",
    volume = "921",
    number = "1",
    pages = "87",
    year = "2021"
}

@article{Lazarian:1998wd,
    author = "Lazarian, A. and Vishniac, E. T.",
    title = "{Reconnection in a weakly stochastic field}",
    eprint = "astro-ph/9811037",
    archivePrefix = "arXiv",
    reportNumber = "POPE-783",
    doi = "10.1086/307233",
    journal = "Astrophys. J.",
    volume = "517",
    pages = "700--718",
    year = "1999"
}

@ARTICLE{Melrose_1969,
       author = {{Melrose}, Donald B.},
        title = "{On the Formation of Energy Spectra in Synchrotron Sources}",
      journal = {Astrophys. Space Sci.},
         year = 1969,
        month = oct,
       volume = {5},
       number = {2},
        pages = {131-149},
          doi = {10.1007/BF00650288},
       adsurl = {https://ui.adsabs.harvard.edu/abs/1969Ap&SS...5..131M}
}

@article{Xu:2017ypi,
    author = "Xu, Siyao and Zhang, Bing",
    title = "{Adiabatic non-resonant acceleration in magnetic turbulence and hard spectra of gamma-ray bursts}",
    eprint = "1708.08029",
    archivePrefix = "arXiv",
    primaryClass = "astro-ph.HE",
    doi = "10.3847/2041-8213/aa88b1",
    journal = "Astrophys. J. Lett.",
    volume = "846",
    number = "2",
    pages = "L28",
    year = "2017"
}

@article{Guo:2019acp,
    author = "Guo, Fan and Li, Xiaocan and Daughton, William and Kilian, Patrick and Li, Hui and Liu, Yi-Hsin and Yan, Wangcheng and Ma, Dylan",
    title = "{Determining the Dominant Acceleration Mechanism during Relativistic Magnetic Reconnection in Large-scale Systems}",
    eprint = "1901.08308",
    archivePrefix = "arXiv",
    primaryClass = "astro-ph.HE",
    doi = "10.3847/2041-8213/ab2a15",
    journal = "Astrophys. J. Lett.",
    volume = "879",
    number = "2",
    pages = "L23",
    year = "2019"
}

@BOOK{Dermer_2009,
       author = {{Dermer}, Charles D. and {Menon}, Govind},
        title = "{High Energy Radiation from Black Holes: Gamma Rays, Cosmic Rays, and Neutrinos}",
         year = 2009,
       adsurl = {https://ui.adsabs.harvard.edu/abs/2009herb.book.....D}
}

@article{Guo:2016yfq,
    author = "Guo, Fan and Li, Hui and Daughton, William and Li, Xiaocan and Liu, Yi-Hsin",
    title = "{Particle Acceleration during Magnetic Reconnection in a Low-beta Pair Plasma}",
    eprint = "1604.02924",
    archivePrefix = "arXiv",
    primaryClass = "astro-ph.HE",
    doi = "10.1063/1.4948284",
    journal = "Phys. Plasmas",
    volume = "23",
    pages = "055708",
    year = "2016"
}

@article{Matthews:2020lig,
    author = "Matthews, James and Bell, Anthony and Blundell, Katherine",
    title = "{Particle acceleration in astrophysical jets}",
    eprint = "2003.06587",
    archivePrefix = "arXiv",
    primaryClass = "astro-ph.HE",
    doi = "10.1016/j.newar.2020.101543",
    journal = "New Astron. Rev.",
    volume = "89",
    pages = "101543",
    year = "2020"
}

@article{Lemoine:2023sxw,
    author = "Lemoine, Martin",
    title = "{Particle transport through localized interactions with sharp magnetic field bends in MHD turbulence}",
    eprint = "2304.03023",
    archivePrefix = "arXiv",
    primaryClass = "physics.plasm-ph",
    doi = "10.1017/S0022377823000946",
    journal = "J. Plasma Phys.",
    volume = "89",
    number = "5",
    pages = "175890501",
    year = "2023"
}

@ARTICLE{Huang_2024,
       author = {{Huang}, Rui and {Howes}, Gregory G. and {McCubbin}, Andrew J.},
        title = "{The velocity-space signature of transit-time damping}",
      journal = {Journal of Plasma Physics},
         year = 2024,
        month = sep,
       volume = {90},
       number = {4},
          eid = {535900401},
        pages = {535900401},
          doi = {10.1017/S0022377824000667},
archivePrefix = {arXiv},
       eprint = {2401.16697},
 primaryClass = {physics.plasm-ph},
       adsurl = {https://ui.adsabs.harvard.edu/abs/2024JPlPh..90d5301H}
}

@article{Yang:2018zrd,
    author = "Yang, Yuan-Pei and Zhang, Bing",
    title = "{Synchrotron radiation from electrons with a pitch-angle distribution}",
    eprint = "1808.05170",
    archivePrefix = "arXiv",
    primaryClass = "astro-ph.HE",
    doi = "10.3847/2041-8213/aada4f",
    journal = "Astrophys. J. Lett.",
    volume = "864",
    number = "1",
    pages = "L16",
    year = "2018"
}

@article{Das:2025vqd,
    author = "Das, Saikat and Zhang, Bing and Razzaque, Soebur and Xu, Siyao",
    title = "{Cosmic-Ray Constraints on the Flux of Ultrahigh-energy Neutrino Event KM3-230213A}",
    eprint = "2504.10847",
    archivePrefix = "arXiv",
    primaryClass = "astro-ph.HE",
    doi = "10.3847/1538-4357/adf8de",
    journal = "Astrophys. J.",
    volume = "991",
    number = "1",
    pages = "96",
    year = "2025"
}

@BOOK{Schlickeiser_2002,
       author = {{Schlickeiser}, Reinhard},
        title = "{Cosmic Ray Astrophysics}",
         year = 2002,
       adsurl = {https://ui.adsabs.harvard.edu/abs/2002cra..book.....S}
}

@article{Cho:2005mb,
    author = "Cho, Jungyeon and Lazarian, A.",
    title = "{Particle acceleration by mhd turbulence}",
    eprint = "astro-ph/0509385",
    archivePrefix = "arXiv",
    doi = "10.1086/498967",
    journal = "Astrophys. J.",
    volume = "638",
    pages = "811--826",
    year = "2006"
}

@article{KM3NeT:2025npi,
    author = "Aiello, S. and others",
    collaboration = "KM3NeT",
    title = "{Observation of an ultra-high-energy cosmic neutrino with KM3NeT}",
    doi = "10.1038/s41586-024-08543-1",
    journal = "Nature",
    volume = "638",
    number = "8050",
    pages = "376--382",
    year = "2025",
    note = "[Erratum: Nature 640, E3 (2025)]"
}

@article{IceCube:2022der,
    author = "Abbasi, R. and others",
    collaboration = "IceCube",
    title = "{Evidence for neutrino emission from the nearby active galaxy NGC 1068}",
    eprint = "2211.09972",
    archivePrefix = "arXiv",
    primaryClass = "astro-ph.HE",
    doi = "10.1126/science.abg3395",
    journal = "Science",
    volume = "378",
    number = "6619",
    pages = "538--543",
    year = "2022"
}

@ARTICLE{LHAASO_2024,
       author = {{LHAASO Collaboration}},
        title = "{Ultrahigh-Energy Gamma-ray Emission Associated with Black Hole-Jet Systems}",
      journal = {arXiv e-prints},
         year = 2024,
        month = oct,
          eid = {arXiv:2410.08988},
        pages = {arXiv:2410.08988},
          doi = {10.48550/arXiv.2410.08988},
archivePrefix = {arXiv},
       eprint = {2410.08988},
 primaryClass = {astro-ph.HE},
       adsurl = {https://ui.adsabs.harvard.edu/abs/2024arXiv241008988L}
}

@article{Lemoine:2024roa,
    author = "Lemoine, M. and Rieger, F.",
    title = "{Neutrinos from stochastic acceleration in black hole environments}",
    eprint = "2412.01457",
    archivePrefix = "arXiv",
    primaryClass = "astro-ph.HE",
    doi = "10.1051/0004-6361/202453296",
    journal = "Astron. Astrophys.",
    volume = "697",
    pages = "A124",
    year = "2025"
}

@article{LHAASO:2023gne,
    author = "{LHAASO Collaboration}",
    collaboration = "LHAASO",
    title = "{Measurement of Ultra-High-Energy Diffuse Gamma-Ray Emission of the Galactic Plane from 10~TeV to 1~PeV with LHAASO-KM2A}",
    eprint = "2305.05372",
    archivePrefix = "arXiv",
    primaryClass = "astro-ph.HE",
    doi = "10.1103/PhysRevLett.131.151001",
    journal = "Phys. Rev. Lett.",
    volume = "131",
    number = "15",
    pages = "151001",
    year = "2023"
}

@article{Xu:2013ppa,
    author = "Xu, Siyao and Yan, Huirong",
    title = "{Cosmic Ray Parallel and Perpendicular Transport in Turbulent Magnetic Fields}",
    eprint = "1307.1346",
    archivePrefix = "arXiv",
    primaryClass = "astro-ph.HE",
    doi = "10.1088/0004-637X/779/2/140",
    journal = "Astrophys. J.",
    volume = "779",
    pages = "140",
    year = "2013"
}

@article{Lazarian:2013dba,
    author = "Lazarian, A. and Yan, Huirong",
    title = "{Superdiffusion of Cosmic Rays: Implications for Cosmic Ray Acceleration}",
    eprint = "1308.3244",
    archivePrefix = "arXiv",
    primaryClass = "astro-ph.HE",
    doi = "10.1088/0004-637X/784/1/38",
    journal = "Astrophys. J.",
    volume = "784",
    pages = "38",
    year = "2014"
}

@article{Pezzi:2021bvc,
    author = "Pezzi, Oreste and Blasi, Pasquale and Matthaeus, William H.",
    title = "{Relativistic Particle Transport and Acceleration in Structured Plasma Turbulence}",
    eprint = "2112.09555",
    archivePrefix = "arXiv",
    primaryClass = "astro-ph.HE",
    doi = "10.3847/1538-4357/ac5332",
    journal = "Astrophys. J.",
    volume = "928",
    number = "1",
    pages = "25",
    year = "2022"
}

@BOOK{Longair_2011,
       author = {{Longair}, Malcolm S.},
        title = "{High Energy Astrophysics}",
         year = 2011,
       adsurl = {https://ui.adsabs.harvard.edu/abs/2011hea..book.....L}
}

@article{Demidem:2019jzn,
    author = "Demidem, Camilia and Lemoine, Martin and Casse, Fabien",
    title = "{Particle acceleration in relativistic turbulence: a theoretical appraisal}",
    eprint = "1909.12885",
    archivePrefix = "arXiv",
    primaryClass = "astro-ph.HE",
    reportNumber = "NORDITA 2020-049",
    doi = "10.1103/PhysRevD.102.023003",
    journal = "Phys. Rev. D",
    volume = "102",
    number = "2",
    pages = "023003",
    year = "2020"
}

@ARTICLE{Sonnerup_1969,
       author = {{Sonnerup}, B.~U. {\"O}.},
        title = "{Acceleration of particles reflected at a shock front}",
      journal = {J. Geophys. Res.},
         year = 1969,
        month = jan,
       volume = {74},
       number = {5},
        pages = {1301},
          doi = {10.1029/JA074i005p01301},
       adsurl = {https://ui.adsabs.harvard.edu/abs/1969JGR....74.1301S}
}

@ARTICLE{Xu_2023,
       author = {{Xu}, Siyao and {Lazarian}, Alex},
        title = "{Turbulent Reconnection Acceleration}",
      journal = {Astrophys. J.},
         year = 2023,
        month = jan,
       volume = {942},
       number = {1},
          eid = {21},
        pages = {21},
          doi = {10.3847/1538-4357/aca32c},
archivePrefix = {arXiv},
       eprint = {2211.08444},
 primaryClass = {physics.plasm-ph},
       adsurl = {https://ui.adsabs.harvard.edu/abs/2023ApJ...942...21X}
}

@ARTICLE{Xu_2022,
       author = {{Xu}, Siyao and {Lazarian}, Alex},
        title = "{Shock Acceleration with Oblique and Turbulent Magnetic Fields}",
      journal = {Astrophys. J.},
         year = 2022,
        month = jan,
       volume = {925},
       number = {1},
          eid = {48},
        pages = {48},
          doi = {10.3847/1538-4357/ac3824},
archivePrefix = {arXiv},
       eprint = {2111.04759},
 primaryClass = {astro-ph.HE},
       adsurl = {https://ui.adsabs.harvard.edu/abs/2022ApJ...925...48X}
}

@ARTICLE{Fisk_1976,
       author = {{Fisk}, L.~A.},
        title = "{The acceleration of energetic particles in the interplanetary medium by transit time damping}",
      journal = {J. Geophys. Res.},
         year = 1976,
        month = sep,
       volume = {81},
       number = {25},
        pages = {4633},
          doi = {10.1029/JA081i025p04633},
       adsurl = {https://ui.adsabs.harvard.edu/abs/1976JGR....81.4633F}
}

@ARTICLE{Hu_2022,
       author = {{Hu}, Yue and {Lazarian}, A. and {Xu}, Siyao},
        title = "{Superdiffusion of cosmic rays in compressible magnetized turbulence}",
      journal = {Mon. Not. R. Astron. Soc.},
         year = 2022,
        month = may,
       volume = {512},
       number = {2},
        pages = {2111-2124},
          doi = {10.1093/mnras/stac319},
archivePrefix = {arXiv},
       eprint = {2111.15066},
 primaryClass = {astro-ph.GA},
       adsurl = {https://ui.adsabs.harvard.edu/abs/2022MNRAS.512.2111H}
}

@article{Vega:2024pkx,
    author = "Vega, Cristian and Boldyrev, Stanislav and Roytershteyn, Vadim",
    title = "{Particle Acceleration in Relativistic Alfv{\'e}nic Turbulence}",
    eprint = "2405.07891",
    archivePrefix = "arXiv",
    primaryClass = "physics.plasm-ph",
    doi = "10.3847/1538-4357/ad5f8f",
    journal = "Astrophys. J.",
    volume = "971",
    number = "1",
    pages = "106",
    year = "2024"
}

@book{oppenheim99,
  author = {Oppenheim, Alan V. and Schafer, Ronald W. and Buck, John R.},
  edition = {Second},
  publisher = {Prentice-hall Englewood Cliffs},
  title = {Discrete-Time Signal Processing},
  year = 1999
}

@article{Takeuchi_2002,
  title = {Relativistic $E\ifmmode\times\else\texttimes\fi{}B$ acceleration},
  author = {Takeuchi, Satoshi},
  journal = {Phys. Rev. E},
  volume = {66},
  issue = {3},
  pages = {037402},
  numpages = {4},
  year = {2002},
  month = {Sep},
  publisher = {American Physical Society},
  doi = {10.1103/PhysRevE.66.037402},
  url = {https://link.aps.org/doi/10.1103/PhysRevE.66.037402}
}

@ARTICLE{Drury_1983,
       author = {{Drury}, L. Oc.},
        title = "{REVIEW ARTICLE: An introduction to the theory of diffusive shock acceleration of energetic particles in tenuous plasmas}",
      journal = {Reports on Progress in Physics},
         year = 1983,
        month = aug,
       volume = {46},
       number = {8},
        pages = {973-1027},
          doi = {10.1088/0034-4885/46/8/002},
       adsurl = {https://ui.adsabs.harvard.edu/abs/1983RPPh...46..973D}
}

@ARTICLE{Kirk_1989,
       author = {{Kirk}, J.~G. and {Heavens}, A.~F.},
        title = "{Particle acceleration at oblique shock fronts}",
      journal = {Mon. Not. R. Astron. Soc.},
         year = 1989,
        month = aug,
       volume = {239},
        pages = {995-1011},
          doi = {10.1093/mnras/239.3.995},
       adsurl = {https://ui.adsabs.harvard.edu/abs/1989MNRAS.239..995K}
}

@ARTICLE{Lichko_2017,
       author = {{Lichko}, E. and {Egedal}, J. and {Daughton}, W. and {Kasper}, J.},
        title = "{Magnetic Pumping as a Source of Particle Heating and Power-law Distributions in the Solar Wind}",
      journal = {Astrophys. J. Lett.},
         year = 2017,
        month = dec,
       volume = {850},
       number = {2},
          eid = {L28},
        pages = {L28},
          doi = {10.3847/2041-8213/aa9a33},
archivePrefix = {arXiv},
       eprint = {1710.02106},
 primaryClass = {physics.plasm-ph},
       adsurl = {https://ui.adsabs.harvard.edu/abs/2017ApJ...850L..28L}
}

@ARTICLE{Malkov_2026,
       author = {{Malkov}, Mikhail and {Jebaraj}, Immanuel},
        title = "{Magnetic Pumping: Plasma Heating to Particle Acceleration}",
      journal = {arXiv e-prints},
         year = 2026,
        month = jan,
          eid = {arXiv:2601.09807},
        pages = {arXiv:2601.09807},
          doi = {10.48550/arXiv.2601.09807},
archivePrefix = {arXiv},
       eprint = {2601.09807},
 primaryClass = {astro-ph.HE},
       adsurl = {https://ui.adsabs.harvard.edu/abs/2026arXiv260109807M}
}

@ARTICLE{Yang_2019,
       author = {{Yang}, Yan and {Wan}, Minping and {Matthaeus}, William H. and {Shi}, Yipeng and {Parashar}, Tulasi N. and {Lu}, Quanming and {Chen}, Shiyi},
        title = "{Role of magnetic field curvature in magnetohydrodynamic turbulence}",
      journal = {Physics of Plasmas},
         year = 2019,
        month = jul,
       volume = {26},
       number = {7},
          eid = {072306},
        pages = {072306},
          doi = {10.1063/1.5099360},
archivePrefix = {arXiv},
       eprint = {1904.08284},
 primaryClass = {physics.space-ph},
       adsurl = {https://ui.adsabs.harvard.edu/abs/2019PhPl...26g2306Y}
}

@article{Takamoto:2017vhf,
    author = "Takamoto, Makoto and Lazarian, Alexandre",
    title = "{Strong coupling of Alfv{\'e}n and fast modes in compressible relativistic magnetohydrodynamic turbulence in magnetically dominated plasmas}",
    eprint = "1709.00785",
    archivePrefix = "arXiv",
    primaryClass = "astro-ph.HE",
    doi = "10.1093/mnras/stx2292",
    journal = "Mon. Not. Roy. Astron. Soc.",
    volume = "472",
    number = "4",
    pages = "4542--4550",
    year = "2017"
}

@article{Takamoto:2016kdu,
    author = "Takamoto, Makoto and Lazarian, Alexandre",
    title = "{Compressible Relativistic Magnetohydrodynamic Turbulence in Magnetically-Dominated Plasmas And Implications for A Strong-Coupling Regime}",
    eprint = "1610.01373",
    archivePrefix = "arXiv",
    primaryClass = "astro-ph.HE",
    doi = "10.3847/2041-8205/831/2/L11",
    journal = "Astrophys. J. Lett.",
    volume = "831",
    number = "2",
    pages = "L11",
    year = "2016"
}

@article{Sebastian:2025dex,
    author = "Sebastian, Samuel T. and Comisso, Luca",
    title = "{Magnetic Field-line Curvature and Its Role in Particle Acceleration by Magnetically Dominated Turbulence}",
    eprint = "2510.20628",
    archivePrefix = "arXiv",
    primaryClass = "astro-ph.HE",
    doi = "10.3847/2041-8213/ae1696",
    journal = "Astrophys. J. Lett.",
    volume = "994",
    number = "1",
    pages = "L1",
    year = "2025"
}

@BOOK{Ginzburg_1964,
       author = {{Ginzburg}, V.~L. and {Syrovatskii}, S.~I.},
        title = "{The Origin of Cosmic Rays}",
         year = 1964,
       adsurl = {https://ui.adsabs.harvard.edu/abs/1964ocr..book.....G}
}

@article{Keenan:2013rza,
    author = "Keenan, Brett D. and Medvedev, Mikhail V.",
    title = "{Particle transport and radiation production in sub-Larmor-scale electromagnetic turbulence}",
    eprint = "1304.3959",
    archivePrefix = "arXiv",
    primaryClass = "astro-ph.HE",
    doi = "10.1103/PhysRevE.88.013103",
    journal = "Phys. Rev. E",
    volume = "88",
    number = "1",
    pages = "013103",
    year = "2013"
}

@ARTICLE{1988SvAL...14..255P,
       author = {{Ptuskin}, V.~S.},
        title = "{Cosmic-Ray Acceleration by Long-Wave Turbulence}",
      journal = {Soviet Astronomy Letters},
         year = 1988,
        month = mar,
       volume = {14},
        pages = {255},
       adsurl = {https://ui.adsabs.harvard.edu/abs/1988SvAL...14..255P}
}

@article{Lubke:2025day,
    author = {L{\"u}bke, Jeremiah and Effenberger, Frederic and Wilbert, Mike and Fichtner, Horst and Grauer, Rainer},
    title = "{Anisotropic Cosmic Ray Transport in strong MHD Turbulence due to Magnetic Mirroring and Resonant Curvature Scattering}",
    eprint = "2509.15320",
    archivePrefix = "arXiv",
    primaryClass = "astro-ph.HE",
    month = "9",
    year = "2025"
}

@ARTICLE{2025ApJ...994..142H,
       author = {{Hu}, Yue and {Xu}, Siyao and {Lazarian}, Alex and {Stone}, James M. and {Hopkins}, Philip F.},
        title = "{Cosmic-ray Perpendicular Superdiffusion and Parallel Mirror Diffusion in a Partially Ionized and Turbulent Medium}",
      journal = {\apj},
         year = 2025,
        month = dec,
       volume = {994},
       number = {2},
          eid = {142},
        pages = {142},
          doi = {10.3847/1538-4357/ae1127},
archivePrefix = {arXiv},
       eprint = {2505.07421},
 primaryClass = {astro-ph.GA},
       adsurl = {https://ui.adsabs.harvard.edu/abs/2025ApJ...994..142H}
}

@ARTICLE{Lazarian:2023sh,
       author = {{Lazarian}, Alex and {Xu}, Siyao and {Hu}, Yue},
        title = "{Cosmic ray propagation in turbulent magnetic fields}",
      journal = {Frontiers in Astronomy and Space Sciences},
         year = 2023,
        month = may,
       volume = {10},
          eid = {1154760},
        pages = {1154760},
          doi = {10.3389/fspas.2023.1154760},
archivePrefix = {arXiv},
       eprint = {2304.02684},
 primaryClass = {astro-ph.GA},
       adsurl = {https://ui.adsabs.harvard.edu/abs/2023FrASS..1054760L}
}

@ARTICLE{Goldreich:1995sr,
       author = {{Goldreich}, P. and {Sridhar}, S.},
        title = "{Toward a Theory of Interstellar Turbulence. II. Strong Alfvenic Turbulence}",
      journal = {\apj},
         year = 1995,
        month = jan,
       volume = {438},
        pages = {763},
          doi = {10.1086/175121},
       adsurl = {https://ui.adsabs.harvard.edu/abs/1995ApJ...438..763G}
}

@article{Das:2025tfq,
    author = "Das, Saikat and Hazra, Srijita and Gupta, Nayantara",
    title = "{Cosmic Clues from Amaterasu: Blazar-driven Ultrahigh-energy Cosmic Rays?}",
    eprint = "2504.16019",
    archivePrefix = "arXiv",
    primaryClass = "astro-ph.HE",
    doi = "10.3847/2041-8213/ade99f",
    journal = "Astrophys. J. Lett.",
    volume = "988",
    number = "1",
    pages = "L8",
    year = "2025"
}

@ARTICLE{2003MNRAS.345..325C,
       author = {{Cho}, Jungyeon and {Lazarian}, A.},
        title = "{Compressible magnetohydrodynamic turbulence: mode coupling, scaling relations, anisotropy, viscosity-damped regime and astrophysical implications}",
      journal = {\mnras},
         year = 2003,
        month = oct,
       volume = {345},
       number = {12},
        pages = {325-339},
          doi = {10.1046/j.1365-8711.2003.06941.x},
archivePrefix = {arXiv},
       eprint = {astro-ph/0301062},
 primaryClass = {astro-ph},
       adsurl = {https://ui.adsabs.harvard.edu/abs/2003MNRAS.345..325C}
}

@article{Chandran:2003xb,
    author = "Chandran, Benjamin D. G. and Maron, Jason L.",
    title = "{Acceleration of energetic particles by large - scale compressible magnetohydrodynamic turbulence}",
    eprint = "astro-ph/0311332",
    archivePrefix = "arXiv",
    doi = "10.1086/377078",
    journal = "Astrophys. J.",
    volume = "603",
    pages = "23--27",
    year = "2004"
}

@article{Sebastian:2026fwz,
    author = {Sebastian, Samuel T. and Xu, Siyao and Hu, Yue and Comisso, Luca and Das, Saikat and N{\"a}ttil{\"a}, Joonas},
    title = "{Turbulence Mode Decomposition and Anisotropy in Magnetically Dominated Collisionless Plasmas}",
    eprint = "2604.20963",
    archivePrefix = "arXiv",
    primaryClass = "physics.plasm-ph",
    doi = "10.3847/1538-4357/ae6336",
    journal = "Astrophys. J.",
    volume = "1003",
    number = "1",
    pages = "79",
    year = "2026"
}

@article{Hosking:2020wom,
    author = "Hosking, David N. and Schekochihin, Alexander A.",
    title = "{Reconnection-Controlled Decay of Magnetohydrodynamic Turbulence and the Role of Invariants}",
    eprint = "2012.01393",
    archivePrefix = "arXiv",
    primaryClass = "physics.flu-dyn",
    doi = "10.1103/PhysRevX.11.041005",
    journal = "Phys. Rev. X",
    volume = "11",
    number = "4",
    pages = "041005",
    year = "2021"
}

@article{Dong:2022crn,
    author = "Dong, Chuanfei and Wang, Liang and Huang, Yi-Min and Comisso, Luca and Sandstrom, Timothy A. and Bhattacharjee, Amitava",
    title = "{Reconnection-driven energy cascade in magnetohydrodynamic turbulence}",
    eprint = "2210.10736",
    archivePrefix = "arXiv",
    primaryClass = "astro-ph.SR",
    doi = "10.1126/sciadv.abn7627",
    journal = "Sci. Adv.",
    volume = "8",
    number = "49",
    pages = "abn7627",
    year = "2022"
}

@ARTICLE{2022Natur.611..677L,
       author = {{Liodakis}, Ioannis and {Marscher}, Alan P. and {Agudo}, Iv{\'a}n and {Berdyugin}, Andrei V. and {Bernardos}, Maria I. and {Bonnoli}, Giacomo and {Borman}, George A. and {Casadio}, Carolina and {Casanova}, V{\'\i}ctor and {Cavazzuti}, Elisabetta and {Rodriguez Cavero}, Nicole and {Di Gesu}, Laura and {Di Lalla}, Niccol{\'o} and {Donnarumma}, Immacolata and {Ehlert}, Steven R. and {Errando}, Manel and {Escudero}, Juan and {Garc{\'\i}a-Comas}, Maya and {Ag{\'\i}s-Gonz{\'a}lez}, Beatriz and {Husillos}, C{\'e}sar and {Jormanainen}, Jenni and {Jorstad}, Svetlana G. and {Kagitani}, Masato and {Kopatskaya}, Evgenia N. and {Kravtsov}, Vadim and {Krawczynski}, Henric and {Lindfors}, Elina and {Larionova}, Elena G. and {Madejski}, Grzegorz M. and {Marin}, Fr{\'e}d{\'e}ric and {Marchini}, Alessandro and {Marshall}, Herman L. and {Morozova}, Daria A. and {Massaro}, Francesco and {Masiero}, Joseph R. and {Mawet}, Dimitri and {Middei}, Riccardo and {Millar-Blanchaer}, Maxwell A. and {Myserlis}, Ioannis and {Negro}, Michela and {Nilsson}, Kari and {O'Dell}, Stephen L. and {Omodei}, Nicola and {Pacciani}, Luigi and {Paggi}, Alessandro and {Panopoulou}, Georgia V. and {Peirson}, Abel L. and {Perri}, Matteo and {Petrucci}, Pierre-Olivier and {Poutanen}, Juri and {Puccetti}, Simonetta and {Romani}, Roger W. and {Sakanoi}, Takeshi and {Savchenko}, Sergey S. and {Sota}, Alfredo and {Tavecchio}, Fabrizio and {Tinyanont}, Samaporn and {Vasilyev}, Andrey A. and {Weaver}, Zachary R. and {Zhovtan}, Alexey V. and {Antonelli}, Lucio A. and {Bachetti}, Matteo and {Baldini}, Luca and {Baumgartner}, Wayne H. and {Bellazzini}, Ronaldo and {Bianchi}, Stefano and {Bongiorno}, Stephen D. and {Bonino}, Raffaella and {Brez}, Alessandro and {Bucciantini}, Niccol{\'o} and {Capitanio}, Fiamma and {Castellano}, Simone and {Ciprini}, Stefano and {Costa}, Enrico and {De Rosa}, Alessandra and {Del Monte}, Ettore and {Di Marco}, Alessandro and {Doroshenko}, Victor and {Dov{\v{c}}iak}, Michal and {Enoto}, Teruaki and {Evangelista}, Yuri and {Fabiani}, Sergio and {Ferrazzoli}, Riccardo and {Garcia}, Javier A. and {Gunji}, Shuichi and {Hayashida}, Kiyoshi and {Heyl}, Jeremy and {Iwakiri}, Wataru and {Karas}, Vladimir and {Kitaguchi}, Takao and {Kolodziejczak}, Jeffery J. and {La Monaca}, Fabio and {Latronico}, Luca and {Maldera}, Simone and {Manfreda}, Alberto and {Marinucci}, Andrea and {Matt}, Giorgio and {Mitsuishi}, Ikuyuki and {Mizuno}, Tsunefumi and {Muleri}, Fabio and {Ng}, Stephen C.-Y. and {Oppedisano}, Chiara and {Papitto}, Alessandro and {Pavlov}, George G. and {Pesce-Rollins}, Melissa and {Pilia}, Maura and {Possenti}, Andrea and {Ramsey}, Brian D. and {Rankin}, John and {Ratheesh}, Ajay and {Sgr{\'o}}, Carmelo and {Slane}, Patrick and {Soffitta}, Paolo and {Spandre}, Gloria and {Tamagawa}, Toru and {Taverna}, Roberto and {Tawara}, Yuzuru and {Tennant}, Allyn F. and {Thomas}, Nicolas E. and {Tombesi}, Francesco and {Trois}, Alessio and {Tsygankov}, Sergey and {Turolla}, Roberto and {Vink}, Jacco and {Weisskopf}, Martin C. and {Wu}, Kinwah and {Xie}, Fei and {Zane}, Silvia},
        title = "{Polarized blazar X-rays imply particle acceleration in shocks}",
      journal = {\nat},
         year = 2022,
        month = nov,
       volume = {611},
       number = {7937},
        pages = {677-681},
          doi = {10.1038/s41586-022-05338-0},
archivePrefix = {arXiv},
       eprint = {2209.06227},
 primaryClass = {astro-ph.HE},
       adsurl = {https://ui.adsabs.harvard.edu/abs/2022Natur.611..677L}
}

@article{MAGIC:2024tyq,
    author = "Abe, S. and others",
    collaboration = "MAGIC",
    title = "{Insights into the broadband emission of the TeV blazar Mrk 501 during the first X-ray polarization measurements}",
    eprint = "2401.08560",
    archivePrefix = "arXiv",
    primaryClass = "astro-ph.HE",
    doi = "10.1051/0004-6361/202348709",
    journal = "Astron. Astrophys.",
    volume = "685",
    pages = "A117",
    year = "2024"
}

@article{Guo:2025zso,
    author = "Guo, Fan and French, Omar and Zhang, Qile and Li, Xiaocan and Seo, Jeongbhin",
    title = "{Particle Injection Problem in Magnetic Reconnection and Turbulence}",
    eprint = "2506.19938",
    archivePrefix = "arXiv",
    primaryClass = "physics.plasm-ph",
    doi = "10.1007/s11214-025-01226-x",
    journal = "Space Sci. Rev.",
    volume = "221",
    number = "8",
    pages = "103",
    year = "2025"
}
\bibliographystyle{aasjournal}



\end{document}